\documentclass[reprint,aps,prc,floatfix,nofootinbib,superscriptaddress]{revtex4-2} 

\pdfoutput=1
\usepackage[labelfont={small},subrefformat=parens,caption=false]{subfig}
\usepackage{amssymb}
\usepackage[utf8]{inputenc}
\usepackage[T1]{fontenc}
\usepackage{natbib}

\usepackage{mathtools} 
\usepackage{mleftright} \mleftright 
\usepackage[dvipsnames]{xcolor}
\usepackage{xspace}
\usepackage{bm} 
\usepackage{braket}
\usepackage[safe]{tipa}
\usepackage{multirow}
\usepackage{array} 
\usepackage{dcolumn} 
\usepackage{booktabs} 
\usepackage{hyperref}
\hypersetup{
  colorlinks   = true,
  urlcolor     = blue, 
  linkcolor    = blue, 
  citecolor   = blue 
}
\usepackage[nameinlink]{cleveref}

\newcommand{\leftd}{\left.\kern-\nulldelimiterspace}
\newcommand{\rightd}{\right.\kern-\nulldelimiterspace}
\newcolumntype{C}[1]{>{\centering\arraybackslash}p{#1}} 
\newcolumntype{n}{>{$}c<{$}} 
\newcolumntype{N}{>{$}r<{$}} 
\newcommand{\rcite}[1]{Ref.~\cite{#1}}
\newcommand{\rscite}[1]{Refs.~\cite{#1}}
\newcommand{\unit}[1]{\ \text{#1}}

\newcommand{\dd}{\mathrm{d}} 
\newcommand{\ee}{\mathrm{e}} 
\DeclareMathOperator{\Tr}{Tr}
\newcommand{\vnabla}{\bm{\nabla}}
\renewcommand{\vr}{\mathbf{r}}
\newcommand{\vR}{\mathbf{R}}
\newcommand{\vx}{\mathbf{x}}
\newcommand{\vJ}{\mathbf{J}}
\newcommand{\obdm}{\bm{\rho}}
\newcommand{\Ekin}{E_\text{kin}}
\newcommand{\EH}{E_\text{H}^\chi}
\newcommand{\EF}{E_\text{F}^\chi}
\newcommand{\ESk}{E_\text{Skyrme}}
\newcommand{\ECou}{E_\text{Coulomb}}
\newcommand{\Epair}{E_\text{pair}}
\newcommand{\SatDens}{\rho_c} 
\newcommand{\SatEn}{E_\text{sat}} 
\newcommand{\Incomp}{K} 
\newcommand{\SymEn}{a_\text{sym}} 
\newcommand{\SlopePar}{L_\text{sym}} 
\newcommand{\ScEffMassInv}{M_\text{s}^{\ast -1}} 
\newcommand{\VecEffMassInv}{M_\text{v}^{\ast -1}} 
\newcommand{\ChargeRadius}{r_\text{ch}} 
\newcommand{\GSEnergy}{E} 
 
\newcommand{\NSepEnergy}{S_\text{2n}} 
\newcommand{\PSepEnergy}{S_\text{2p}} 
\newcommand{\elm}[2]{$^{#2}$#1} 
\newcommand{\NNLO}{N$^2$LO\xspace}
\newcommand{\NLOD}{NLO$\Delta$\xspace}
\newcommand{\NNLOD}{N$^2$LO$\Delta$\xspace}
\newcommand{\NLODThreeN}{NLO$\Delta$+3N\xspace}
\newcommand{\NNLOThreeN}{N$^2$LO+3N\xspace}
\newcommand{\NNLODThreeN}{N$^2$LO$\Delta$+3N\xspace}
\newcommand{\NNNLO}{N$^3$LO\xspace}
\newcommand{\nochi}{no chiral\xspace}
\newcommand{\minchi}{min.\ chiral\xspace}
\newcommand{\gude}{GUDE\xspace}
\newcommand{\classesZeroOne}{classes 0 and~1\xspace}
\newcommand{\classZero}{class~0\xspace}
\newcommand{\classOne}{class~1\xspace}
\newcommand{\classTwo}{class~2\xspace}
\newcommand{\classTwoHyphen}{class-2\xspace}
\newcommand{\PenaltyFunc}{loss function\xspace}

\crefname{table}{Table}{Tables}
\crefname{figure}{Fig.}{Figs.}
\Crefname{figure}{Figure}{Figures}
\crefname{equation}{Eq.}{Eqs.}
\Crefname{equation}{Equation}{Equations}
\crefname{section}{Sec.}{Secs.}
\Crefname{section}{Section}{Sections}

\begin{document}

\title{Optimized nuclear energy density functionals including long-range pion contributions}

\author{L.~Zurek}
\email{lzurek@theorie.ikp.physik.tu-darmstadt.de}
\affiliation{Technische Universit\"at Darmstadt, Department of Physics, 64289 Darmstadt, Germany}
\affiliation{ExtreMe Matter Institute EMMI, GSI Helmholtzzentrum f\"ur Schwerionenforschung GmbH, 64291 Darmstadt, Germany}

\author{S.~K.~Bogner}
\email{bogner@frib.msu.edu}
\affiliation{Facility for Rare Isotope Beams and Department of Physics and Astronomy,  \\
\mbox{Michigan State University, East Lansing, MI 48824, USA}
}

\author{R.~J.~Furnstahl}
\email{furnstahl.1@osu.edu}
\affiliation{Department of Physics, The Ohio State University, Columbus, OH 43210, USA}

\author{R.~\surname{Navarro~Pérez}}
\email{rnavarroperez@msjc.edu}
\affiliation{Department of Physics, San Diego State University, 5500 Campanile Drive, San Diego, CA 92182-1233, USA}
\affiliation{Department of Physics, Mt.\ San Jacinto College, San Jacinto, CA 92583, USA}

\author{N.~\surname{Schunck}}
\email{schunck1@llnl.gov}
\affiliation{Nuclear and Data Theory group, Nuclear and Chemical Science Division, Lawrence Livermore National Laboratory, Livermore, CA 94550, USA}

\author{A.~Schwenk}
\email{schwenk@physik.tu-darmstadt.de}
\affiliation{Technische Universit\"at Darmstadt, Department of Physics, 64289 Darmstadt, Germany}
\affiliation{ExtreMe Matter Institute EMMI, GSI Helmholtzzentrum f\"ur Schwerionenforschung GmbH, 64291 Darmstadt, Germany}
\affiliation{Max-Planck-Institut für Kernphysik, Saupfercheckweg 1, 69117 Heidelberg, Germany}

\begin{abstract}
Nuclear energy density functionals successfully reproduce properties of nuclei across almost the entire nuclear chart. 
However, nearly all available functionals are phenomenological in nature and lack a rigorous connection to systematically improvable nuclear forces.
This issue might be solved with an energy density functional obtained from first principles.
As an intermediate step towards this goal we construct the \gude family of functionals that is obtained from a hybrid scheme consisting
of long-range pion-exchange contributions derived from chiral effective field theory at the Hartree-Fock level and a phenomenological Skyrme part.
When including pion contributions beyond next-to-leading order in the chiral expansion, we find significant improvements over a reference Skyrme functional constructed following the same protocol. 
We analyze the importance of different pion contributions and identify which terms drive the observed improvements.
Since pions are incorporated without adding further optimization parameters to the functionals, the improvements can be attributed to the functional form of these terms.
Our work therefore suggests that the considered chiral contributions constitute useful ingredients for true ab initio energy density functionals.

\end{abstract}

\maketitle

\section{\label{sec:intro}Introduction}

Tremendous progress has been made in calculating nuclear structure from first principles~\cite{Herg20AbInitio,Hebe21ThreeN}, pushing descriptions toward heavy~\cite{Miya22heavy,Hu22Pb208,Hebe23JacobiNO} 
and doubly open-shell~\cite{Gebr17IMNCSM,Stro17ENO,Miya20MVSIMSRG,Stro21dripline,Soma21DOpenGor,Fros22PGCMPT2} 
nuclei,
and employing high-precision interaction models~\cite{Ente20N5LOScat,Epel20SemiOver,Tews22ForcColl} 
and high-order many-body methods~\cite{Hage14RPP,Raim18ADC33N,Hein21IMSRG3}.
However, due to their huge numerical cost, these microscopic approaches, usually generically referred to as ab initio methods~\cite{Ekst23AbInitio}, are not yet ready to be employed in large-scale, high-precision calculations of nuclear ground-state observables.
Even if one could overcome this computational challenge,
it is unclear whether ab initio calculations are going to be able to compete with less microscopic methods regarding the accuracy they can achieve.
At present, they generally cannot~\cite{Drut10PPNP,Hage14RPP,Herg20AbInitio,Hebe21ThreeN}.

Nuclear density functional theory (DFT) ~\cite{Bend03RMP,Schu19EDFBook} 
is currently the most microscopic theoretical framework that
can be used in global surveys thanks to its favorable computational scaling~\cite{Dugu22PGCMEDF}.
It is rooted in the seminal work by Hohenberg and Kohn proving the existence of a universal functional of the density which, when minimized for fixed particle number, gives the ground state density and energy of a many-body system confined in an external potential~\cite{Hohe64HKtheorm}. 
While this is most commonly employed for the description of electronic systems, later works extended the existence proof to self-bound systems as constituted by finite nuclei~\cite{engel2007intrinsicdensity,barnea2007density,giraud2008existence,messud2009density}.
In practice most calculations are carried out in the Kohn-Sham formulation of DFT~\cite{Kohn65KS}, which allows for an efficient description of the kinetic energy of the system and of shell effects 
by expressing the density of interest in terms of auxiliary single-particle orbitals of an independent-particle system.

In nuclear physics different ansatze have been established for the form of the energy density functional (EDF).
In the nonrelativistic sector, the Skyrme~\cite{Ston07Skyrme} and Gogny~\cite{Robl19Gogny} EDFs are based on effective nucleon-nucleon interactions.
Genuine energy functionals (not derived from an underlying potential) include the Fayans~\cite{Rein17FayEDF}, the SeaLL1~\cite{Bulg18SeaLL1}, and the BCPM~\cite{Bald17BCPMStar} functionals.
Different forms are also available in covariant DFT;
see, e.g., \rscite{Niks11RelaEDF,Pere21RadiiCov}.
Here we will limit ourselves to nonrelativistic functionals.

Significant progress in nuclear DFT has been achieved by using increasingly sophisticated parameter optimization protocols but it is widely believed that this avenue has been explored to such a degree that further improvements, necessary for instance for the description of r-process nucleosynthesis~\cite{Pano10ReacRate,Mart16massrpro,Zhu21KiloDep,Spro20rProSkyr,Gori23NuclRPro} or of single-particle
energies~\cite{Kort08SkyrmeSP}, need to come from elsewhere~\cite{Kort08SkyrmeSP,Kort14UNEDF2,McDo15UncQuaEDF}.
The two most obvious routes are the explicit treatment of static correlations within a multireference framework~\cite{Egid16BMF,Robl19Gogny,Gras19EDFBMF,Schu19EDFBook} and the extension of the form of the employed EDFs.

In the latter direction, different empirical strategies have been pursued (see, e.g., 
\rscite{Carl08EDF,Raim14NonloEDF,Xion16SkyDD,Beck17N2LOSky,Bacz18ISB,Benn20NonloEDF,Bata22Gogny3G}).
They often consist in adding similar or higher-order terms to existing EDF structures and typically involve introducing additional adjustable parameters.
Properly fitting such parameters is a nontrivial task since they cannot always be well constrained with available experimental data. 
This does not address the phenomenological nature of the EDFs, which is the root cause for potentially uncontrolled extrapolations outside the fitting regions~\cite{Erle10SkyrProb,Erle12Nature,McDo15UncQuaEDF,Mump16IndiRPro,Tani20CovDFTrp}.

A unifying construction principle for nuclear EDFs might therefore be helpful.
While different ideas to formulate an effective field theory (EFT) for EDFs have been discussed~\cite{Gras16EFTEDF,Furn20EDFasEFT,Frab23EDFPath1}, none of them has been implemented yet.
Alternatively, one can remain within the overall framework of nuclear DFT while seeking guidance 
from microscopic ab initio theories.
By employing interactions derived from chiral EFT, which establishes a construction scheme based on a power counting estimating the importance of individual terms~\cite{Epel09RMP,Mach11PR}, ab initio calculations become systematically improvable by going to higher orders in the chiral expansion.
At present the most accurate potentials are constructed at fifth order
for nucleon-nucleon (NN) forces~\cite{Ente20N5LOScat,Epel20SemiOver} 
and fourth order for three-nucleon (3N) forces~\cite{Hebe21ThreeN}.
Different ideas exist for how to combine ab initio approaches and nuclear DFT~\cite{DFTRG1,Lesi09NoEmPair,Drut10PPNP,Gamb11EDFfrBHF,DFTRG2,Bald17BCPMStar,Shen19aiCovDFT,Salv20AbInEDF,Burr21YGLONucl,Mari21AbInEDF,Dugu22PGCMEDF}.
They range from determining EDF parameters~\cite{Gamb11EDFfrBHF,Salv20AbInEDF} and constraining the form of some functional terms~\cite{Lesi09NoEmPair,Bald17BCPMStar,Burr21YGLONucl} based on microscopic calculations
to ideas for a full determination of the functional form from a chiral interaction model~\cite{DFTRG1,DFTRG2,Dugu22PGCMEDF}.

In this work, we follow a hybrid strategy first suggested in \rscite{Drut10PPNP,Gebr10DME}.
It consists in adding terms arising from pion exchanges as described by chiral EFT interactions at the Hartree-Fock (HF) level on top of a Skyrme EDF structure.
There are two motivations for this strategy. 
First, the form of the Skyrme EDF corresponds to calculating HF energies from contact interactions.
Following chiral EFT, the first additional degree of freedom that appears when increasing the resolution of the description of the considered systems are the pions exchanged between the nucleons. 
Adding them explicitly should lead to a more accurate description of nuclear properties.
Second, one notices that ab initio calculations with chiral EFT interactions
often build correlations on top of an initial mean-field solution.
In our approach, we employ the same interactions but instead of generating correlations via the many-body method, we adjust the short-range part of the interactions.
This is because the dominant bulk correlations in nuclei, e.g., in expansions around HF, appear to be short range in nature~\cite{Zhan18BHFDME} and 
could therefore be mimicked 
by contact interactions.

This semiphenomenological strategy was implemented in a series of papers, \rscite{Gebr10DME,Gebr10DME3N,Stoi10DMEEDF,Gebr11DME,Dyhd16DME,Nava18DMEEDF}.
While improvements over EDFs without chiral terms were observed, the dependency of the results on the order of the chiral interactions showed large variability and puzzling systematics~\cite{Nava18DMEEDF}.
The goal of the present work is to carefully revisit the construction of EDFs incorporating chiral physics via a density-matrix expansion (DME).
We study in detail the dependence of the results on the order of the employed chiral interaction and identify which terms are crucial to obtain improvement over EDFs without pion-exchange terms.
To perform these investigations we construct a new set of nuclear EDFs which we dub ``Germany-USA DME EDFs'' (\gude \footnote{
\emph{gude} 
\textipa{[gu\textlengthmark d\textschwa]} 
is a common greeting in the Hessian dialect of German that is spoken in Darmstadt, among other places. 
} 
for short).

We begin by laying out the theoretical framework of this study in \cref{sec:method}.
In particular, we discuss the structure of the EDFs including the chiral contributions, the numerical setup used to determine nuclear properties from them, and the parameter optimization protocol. 
In \cref{sec:results} we present the obtained \gude parametrizations and investigate their performance by comparing against experimental data.
In particular, we construct a \gude variant which reproduces the main improvements found in this work by adding only a minimal number of terms arising from pion exchanges.
\Cref{sec:analysis} contains a detailed analysis of the order-by-order behavior of the functionals in the \gude family.
We end by summarizing our findings in \cref{sec:SummaryOutlook}, where we also give an outlook on avenues for future work.

\section{\label{sec:method}Method}

The EDFs we construct in this work can be split into six parts according to
\begin{equation}\label{eq:FullEDF}
    E = \EH + \EF + \ESk + \ECou + \Epair + \Ekin \,.
\end{equation}
They are solved at the Hartree-Fock-Bogoliubov (HFB) level using the code \texttt{HFBTHO}~\cite{Nava17HFBTHOv3}, as detailed in \cref{sec:HFB}.
The conventional part of the EDFs consists of the latter four terms. 
The Skyrme part reads
\begin{align}\label{eq:Sk}
    \ESk = &\sum_{t=0,1} \int \! \dd\vR \,
    \bigl[ C_t^{\rho\rho}(\rho_0) \rho_t^2 
    + C_t^{\rho\tau} \rho_t \tau_t 
        \notag \\
    &\null + C_t^{\rho\Delta\rho} \rho_t \Delta\rho_t
    + C_t^{\rho\nabla J} \rho_t \vnabla \cdot \vJ_t\notag \\
    &\null + C_t^{JJ} J_{t,ab} J_{t,ab}
        \bigr] \,,
\end{align}
where 
\begin{equation}
    C_t^{\rho\rho}(\rho_0) = C_{t0}^{\rho\rho} + C_{tD}^{\rho\rho} \rho_0^\gamma \,,
\end{equation}
and the isospin index $t=0$ ($t=1$) labels isoscalar (isovector) densities.
Summations over spacial indices $a$, $b$ are implied.
In \cref{eq:Sk}, we have suppressed the dependence 
on the position $\vR$
of the \mbox{(quasi)}local densities, for which expressions can be found in \rscite{Bend03RMP, Lesi07SkyrTens}.
Since we only apply our EDFs for calculations of even-even nuclei, 
time-odd densities are not taken into account in the construction.
Skyrme and pairing $\Epair$ (\cref{sec:pairing}) contributions contain the parameters that are adjusted to data as described in \cref{sec:optimization}.
The Coulomb energy is obtained here as in \rscite{Kort10edf, Kort12UNEDF1, Kort14UNEDF2, Nava18DMEEDF}: the Hartree term is calculated exactly using the Gaussian substitution method~\cite{Giro83HFBTriax,Stoi13HFBTHOv2} and the exchange term is calculated with the Slater approximation~\cite{Slat50SimplHF};
see \rcite{Mare22HFBTHOv4} for an assessment of the accuracy of these methods.
The kinetic energy is given by
\begin{equation}
    \Ekin = \int \! \dd\vR \, \frac{\hbar^2}{2m} \tau_0(\vR) \,,
\end{equation}
with $\hbar^2/(2m) = 20.73553 \unit{MeV fm$^2$}$.

In \cref{sec:results} we construct a conventional functional, below labeled as ``\nochi'', that contains only these four terms and serves as a reference functional for comparing the performance of the other EDFs that we construct following the same optimization protocol.
These additionally contain the first two terms in \cref{eq:FullEDF}, $\EH$ and $\EF$, which represent the Hartree and Fock energy from pion exchanges, respectively. 
The expressions for the pion exchanges which enter the definitions of $\EH$ and $\EF$ are taken directly from interactions derived from chiral EFT at different orders; see \cref{sec:interaction}. 
Because the low-energy constants of the pion exchanges are determined from few-body data~\cite{Kreb07Deltas} and are not adjusted in the present work, the additional inclusion of these terms does not lead to an increase in the number of adjustable functional parameters.
See \cref{sec:Hartree,sec:Fock} for details regarding the pion Hartree and Fock terms.

While the structure of the functionals constructed here agrees with the one from \rcite{Nava18DMEEDF}, we introduce several changes and improve various aspects
in the construction and optimization of the functionals compared to that work.
These changes, stated in detail in  \cref{sec:interaction,sec:Hartree,sec:Fock,sec:pairing,sec:HFB,sec:optimization}, 
are mostly driven by the idea to enable a cleaner comparison of the functionals constructed at (different) chiral orders.

\subsection{\label{sec:interaction}Chiral interactions}

For the construction of the EDFs we consider pion exchanges at different orders in the chiral expansion up to next-to-next-to-leading order (\NNLO) both with and without the explicit inclusion of intermediate $\Delta$ isobars as well as with and without three-nucleon (3N) forces.
Chiral EFT interactions contain pion exchanges and contact interactions.
We take only the finite-range parts of the pion exchanges explicitly into account. 
Expressions for the corresponding interaction terms in coordinate space are given in \rscite{Dyhd16DME,Nava18DMEEDF}.
The low-energy constants that appear are taken from the determination of \rcite{Kreb07Deltas} (columns ``$Q^2$, no $\Delta$'' and ``$Q^2$, fit 1'' of Table 1 therein).
Note that we use $g_A = 1.27$ and $h_A = 3g_A/\sqrt{2}$ as chosen in \rcite{Kreb07Deltas}.
The previous implementation~\cite{Nava18DMEEDF} used the Fock coefficient functions derived in \rcite{Dyhd16DME} for which the slightly inconsistent combination of $g_A = 1.29$ with low-energy constants from \rcite{Kreb07Deltas} had been considered.
The finite-range interactions are regularized by multiplying them with the local regulator function 
\begin{equation}\label{eq:regulator}
    f(r) = \left[ 1 - \exp \left( - \frac{r^2}{R_c^2} \right) \right]^n \,,
\end{equation}
where we choose $R_c = 1.0 \unit{fm}$ and $n=6$ (cf.~\cite{Epel15EKMInter}).
Investigating the choice of the regularization scheme is left for future work.

Contact interactions as well as correlations involving pions beyond the HF level are assumed to be effectively captured by the EDFs by adjusting the parameters of $\ESk$ and $\Epair$ to data from finite nuclei.

\subsection{\label{sec:Hartree}Long-range Hartree terms}

The Hartree terms from the pion exchanges are included essentially exactly by evaluating the corresponding integrals.
Since we consider only even-even nuclei, the spin density vanishes due to time-reversal symmetry so that only the central part of the NN interactions contribute:
\begin{equation}\label{eq:Hartree}
    \EH = \frac{1}{2} \sum_{t=0,1} \int \! \dd\vR \dd\vr \, V_t(r) \rho_t \left( \vR + \frac{\vr}{2} \right) \rho_t \left( \vR - \frac{\vr}{2} \right) \,.
\end{equation}
To make use of the capability of \texttt{HFBTHO}
to solve the HFB equations for potentials given by sums of Gaussians~\cite{Nava17HFBTHOv3},
we approximate the central chiral potentials as
\begin{align}\label{eq:VCGaussians}
    V_0(r) = V_C(r) \to \tilde V_C (r) &= \sum_{i=1}^N \left( W_i - \frac{H_i}{2} \right) \ee^{-r^2 / \mu_i^2} \,,\\ 
    \label{eq:WCGaussians}
    V_1(r) = W_C(r) \to \tilde W_C (r) &= -\sum_{i=1}^N \frac{H_i}{2} \ee^{-r^2 / \mu_i^2} \,.
\end{align}
A similar idea was implemented in \rcite{Doba09hfodd}. 
Together with $B_i = M_i = 0$ (which do not contribute here due to time-reversal invariance),
\cref{eq:VCGaussians,eq:WCGaussians} correspond to a Gogny-like interaction,
\begin{equation}\label{eq:Gogny}
    V_G = \sum_{i=1}^N (W_i + B_i P_\sigma - H_i P_\tau - M_i P_\sigma P_\tau) \ee^{-r^2 / \mu_i^2} \,.
\end{equation}
Note that in \cref{eq:VCGaussians,eq:WCGaussians,eq:Gogny} we correct several mistakes
compared to \namecrefs{eq:VCGaussians}~(30) to (33) of \rcite{Nava18DMEEDF}.
The wrong equations in \rcite{Nava18DMEEDF} led to an incorrect implementation of the Hartree terms in the functionals constructed therein.

To reproduce the behavior of the regulator [\cref{eq:regulator}] at the origin, the conditions
\begin{align}\label{eq:LastGaussParameters}
    H_N &= - \sum_{i=1}^{N-1} H_i \,, 
    &W_N &= - \sum_{i=1}^{N-1} W_i
\end{align}
are imposed.
The remaining free parameters $W_i$, $H_i$, $\mu_i$ are obtained by a fitting routine.

As in \rcite{Nava18DMEEDF}, $N=5$ Gaussians are used here as a compromise between accuracy of the approximation and computational requirements for evaluating and storing the resulting integrals \cite{Nava19HartGaus}.
The Gaussians used in \rcite{Nava18DMEEDF} were obtained by simultaneously fitting all 13 parameters for the isoscalar $V_C$ and isovector $W_C$ potentials.
Here, we fit first only the nine parameters for the isoscalar potential $V_C$ since it contributes significantly more to the energy of finite nuclei than its isovector counterpart.
We keep the resulting Gaussian widths $\mu_i$ fixed for the subsequent fitting of the remaining four parameters of the isovector potential $W_C$.
We obtain the parameters of the Gaussians by $\chi^2$ minimizations where the loss functions are given by
\begin{equation}
    \chi^2 = \sum_{r} \left\{ r^2 \left[ \tilde V_t(r) - V_t(r)\right] \right\}^2 \,,
\end{equation}
which are evaluated on an evenly spaced grid from $r=0$ to 8~fm with step width 0.125~fm.
We include the $r^2$ prefactor in the definition of the $\chi^2$ to account for the increased importance of larger $r$ due to the presence of the volume element in the Hartree energy, \cref{eq:Hartree}.
This factor had not been included in the determination of the Gaussian parameters in \rscite{Nava18DMEEDF,Nava19HartGaus}.
We provide the Gaussian parameters obtained in the new fit 
in the Supplemental Material~\cite{GUDESupplemental}.

In \cref{fig:VC,fig:WC} we plot $r^2 [V_t(r) - \tilde V_t(r)]$ including contributions up to including \NNLO in the chiral expansion (without explicitly resolved $\Delta$ excitations).
The new fitting strategy improves the fit of $V_C$ without a significant degradation in fitting $W_C$.
When evaluating the Hartree energy expectation value in \elm{Pb}{208} the difference between the value obtained with the exact and the approximated potential at \NNLO is about 5 MeV (on a total Hartree energy of about 4000 MeV) with the Gaussian parameters obtained in this work. 
This is a significant improvement over the difference of 37 MeV obtained with the Gaussian parameters of \rscite{Nava18DMEEDF,Nava19HartGaus}.
Similar improvements are obtained for the fits of the potentials at 
other 
chiral orders.
For these comparisons the underlying single-particle orbitals were generated from a self-consistent HF calculation with the SLy4 EDF~\cite{Chab98SLy}
using the code \texttt{HOSPHE}~\cite{Carl10hosphe}.

\begin{figure}[tbp]
\includegraphics[width=1\linewidth]{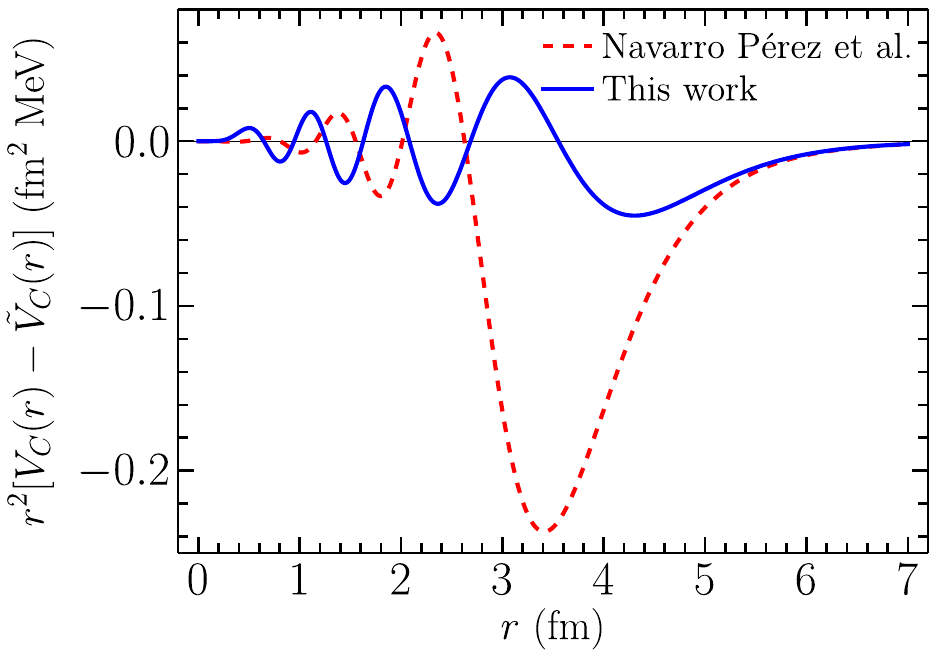}
\caption{\label{fig:VC}
$r^2$-weighted difference between isoscalar central potential at \NNLO in the chiral expansion and its approximations by sums of five Gaussians according to \cref{eq:VCGaussians}.
Both the approximation of \rscite{Nava18DMEEDF,Nava19HartGaus} and the one obtained here are shown.
}
\end{figure} 
\begin{figure}[tbp]
\includegraphics[width=1\linewidth]{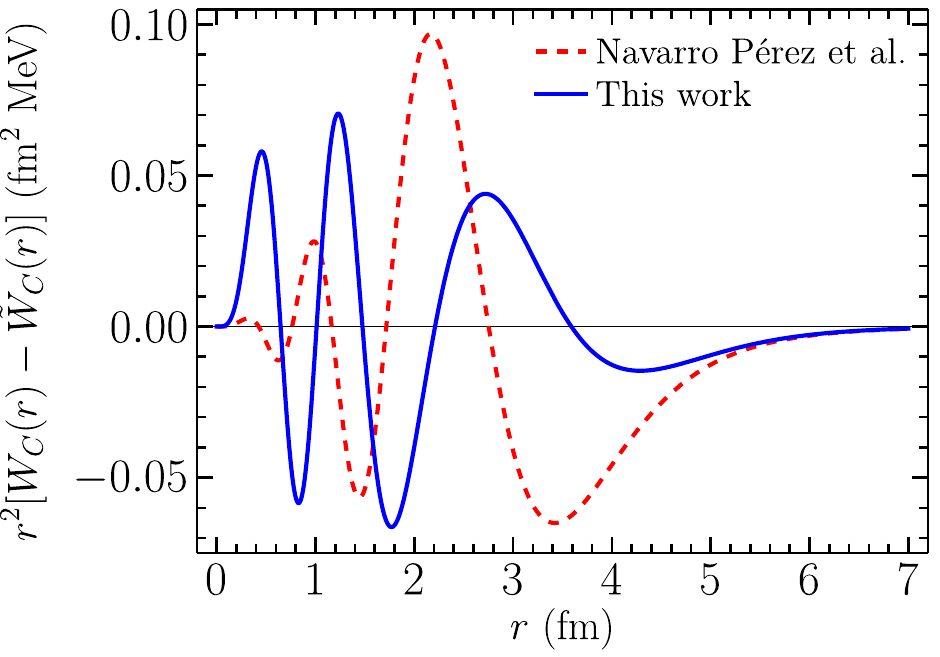}
\caption{\label{fig:WC}
Same as \cref{fig:VC} but for the isovector potential.
}
\end{figure} 

Note that it is not clear if and how the observed improvements translate into improvements of the constructed EDFs.
This is because the Skyrme parameters are fitted to data after adding the terms originating in the chiral potentials and this fitting can (partly) compensate the errors from the nonperfect Gaussian approximations.
For the same reason it is also hard to gauge a priori the impact of other changes we introduced compared to \rcite{Nava18DMEEDF}. 

For later reference we introduce a notation for contributions arising when performing a Taylor expansion of one density entering \cref{eq:Hartree} in the relative coordinate $\vr$ about the argument of the other density.
We write
\begin{equation}\label{eq:HartreeTaylor}
    \EH = \sum_{t=0,1} \int \! \dd\vR \, \sum_{n=0}^\infty T_t^{\rho \Delta^{\hspace{-1pt}n} \hspace{-1pt}\rho} \rho_t(\vR) \Delta^n \rho_t(\vR) \,
\end{equation}
with
\begin{equation}
    T_t^{\rho \Delta^{\hspace{-1pt}n} \hspace{-1pt}\rho} = 2 \pi \int \! \dd r \, r^2 V_t(r) \frac{r^{2n}}{(2n+1)!} \,.
\end{equation}
Finally, we recall that there are no Hartree contributions from the long-range parts of 3N forces at the orders we consider.

\subsection{\label{sec:Fock}Long-range Fock terms}

The Fock energy arising from a local NN interaction $V_\chi$ is given by \begin{align}\label{eq:Fock1}
    E_\text{F,NN}^\chi = &-\frac{1}{2} \Tr_{12}^{\sigma \tau} \int \! \dd\vR \dd\vr \braket{\vr | V_\chi({\bm \sigma_1}, {\bm \sigma_2}, {\bm \tau_1}, {\bm \tau_2})  | \vr } P_{12}^{\sigma\tau} \notag \\
    &\times \obdm^{(1)} \left( \vR - \frac{\vr}{2}, \vR + \frac{\vr}{2} \right) \obdm^{(2)} \left( \vR + \frac{\vr}{2}, \vR - \frac{\vr}{2} \right) \,.
\end{align}
A DME allows one to approximately rewrite the nonlocal one-body density matrix $\obdm$ as a sum of terms in which the nonlocality is factored out~\cite{Zure21DMEs}.
After applying the DME and carrying out the traces and the integral in the nonlocality $\vr$, one obtains a quasilocal approximation for the Fock energy, which for the NN forces used here 
reads~\cite{Dyhd16DME} 
\begin{align}\label{eq:Fock}
    E_\text{F,NN}^\chi = &\sum_{t=0,1} \int \! \dd\vR \,
    \left\{ g_t^{\rho\rho}(\rho_0) \rho_t^2 
    + g_t^{\rho\tau}(\rho_0) \rho_t \tau_t 
    \vphantom{g_t^{JJ,3}} \right.\notag \\
    &\null + g_t^{\rho\Delta\rho}(\rho_0) \rho_t \Delta\rho_t
    + g_t^{JJ,2}(\rho_0) J_{t,ab} J_{t,ab}
    \notag \\
    &\null + \leftd g_t^{JJ,1}(\rho_0) \left[ J_{t,aa} J_{t,bb}
    + J_{t,ab} J_{t,ba} \right] \right\} \,.
\end{align}
As before we consider only terms that contribute in time-reversal invariant systems.
Note $J_{t,aa}=0$ when axial symmetry is conserved~\cite{Stoi05HFBTHOv1}, which is the case for all calculations performed in this work.
\Cref{eq:Fock} looks very similar to the Skyrme part of the functional, \cref{eq:Sk}.
However, in \cref{eq:Fock} the prefactors of the density bilinears (the $g$ coefficient functions $g_t^{u v}$) are not constants 
but functions of the isoscalar density $\rho_0$ and are fixed once one picks a chiral interaction model and a DME variant.

From a computational point of view, using a DME does not provide a significant benefit when considering only chiral NN interactions, but it is a suitable strategy to make the addition of 3N interactions feasible.
For those interactions the equation for the Fock contributions is determined analogously to the NN case and reads~\cite{Dyhd16DME}
\begin{widetext}
\begin{align}
  E_\text{F,3N}^\chi = &\int \! \dd\vR \,
  \left\{
  g^{\rho_0^3}(\rho_0) \rho_0^3 
  + 
  g^{\rho_0^2 \tau_0}(\rho_0) \rho_0^2 \tau_0 
  +
  g^{\rho_0^2 \Delta \rho_0}(\rho_0) \rho_0^2 \Delta \rho_0 
  +
  g^{\rho_0 (\nabla \rho_0)^2}(\rho_0)
  \rho_0 \vnabla \rho_0 \cdot \vnabla \rho_0 
  +
  g^{\rho_0 \rho_1^2}(\rho_0) \rho_0 \rho_1^2 \right.
\notag \\
  & \null + 
  g^{\rho_1^2 \tau_0}(\rho_0) \rho_1^2 \tau_0
  +
  g^{\rho_1^2 \Delta \rho_0}(\rho_0)
  \rho_1^2 \Delta \rho_0
  +
  g^{\rho_0 \rho_1 \tau_1}(\rho_0)
  \rho_0 \rho_1 \tau_1
  +
  g^{\rho_0 \rho_1 \Delta \rho_1}(\rho_0)
  \rho_0 \rho_1 \Delta \rho_1
  +
  g^{\rho_0 (\nabla \rho_1)^2}(\rho_0)
  \rho_0 \vnabla \rho_1 \cdot \vnabla \rho_1
\notag \\
  & \null + 
  \rho_0 \epsilon_{ijk}
  \left[
  g^{\rho_0 \nabla \rho_0 J_0}(\rho_0)
  \nabla_i \rho_0 J_{0,jk}
  +
  g^{\rho_0 \nabla \rho_1 J_1}(\rho_0)
  \nabla_i \rho_1 J_{1,jk}
  \right]
\notag \\
  & \null + 
  \rho_1 \epsilon_{ijk}
  \left[
  g^{\rho_1 \nabla \rho_1 J_0}(\rho_0)
  \nabla_i \rho_1 J_{0,jk}
  +
  g^{\rho_1 \nabla \rho_0 J_1}(\rho_0)
  \nabla_i \rho_0 J_{1,jk}
  \right]
\notag \\
  & \null + 
  \rho_0 \left[
  g^{\rho_0 J_0^2, 1}(\rho_0)
  J_{0,aa} J_{0,bb}
  +
  g^{\rho_0 J_0^2, 2}(\rho_0)
  J_{0,ab} J_{0,ab} 
  +
  g^{\rho_0 J_0^2, 3}(\rho_0)
  J_{0,ab} J_{0,ba}
  \right]
\notag \\
  & \null + 
  \rho_0 \left[
  g^{\rho_0 J_1^2, 1}(\rho_0)
  J_{1,aa} J_{1,bb}
  +
  g^{\rho_0 J_1^2, 2}(\rho_0)
  J_{1,ab} J_{1,ab} 
  +
  g^{\rho_0 J_1^2, 3}(\rho_0)
  J_{1,ab} J_{1,ba}
  \right]
\notag \\
  & \null \left.\vphantom{g^{\rho_0^3}} + 
  \rho_1 \left[
  g^{\rho_1 J_0 J_1, 1}(\rho_0)
  J_{1,aa} J_{0,bb}
  +
  g^{\rho_1 J_0 J_1, 2}(\rho_0)
  J_{1,ab} J_{0,ab} 
  +
  g^{\rho_1 J_0 J_1, 3}(\rho_0)
  J_{1,ab} J_{0,ba}
  \right] \right\}
  \,.
  \label{eq:nnn_edf}
\end{align}
\end{widetext}

In the actual HFB calculations with \texttt{HFBTHO} the $g$ coefficients are approximated with interpolation functions of the form 
\begin{align}
\label{eq:NNarctan}
    g_t^{u v} (\rho_0) &\to \tilde g_t^{u v} (\rho_0) = \tilde g_t^{u v} (0) + \sum_{i=1}^N a_i \arctan\left(b_i \rho_0^{c_i}\right)^i \,, \\
    \label{eq:3Narctan}
    g^{u v w} (\rho_0) &\to \tilde g^{u v w} (\rho_0) = \tilde g^{u v w} (0) + \sum_{i=1}^N a_i \arctan\left(b_i \rho_0^{c_i}\right)^i \,,
\end{align}
where $N=3$ and the coefficients $\tilde g_t^{u v (w)} (0), a_i, b_i, c_i$ are fitted separately for each $g$ coefficient.
For details on the interpolation see \rcite{Nava18DMEEDF}.
Note that Eq.~(47) therein contains an error which is corrected in \cref{eq:NNarctan,eq:3Narctan} above.

In this work we stick to the choice of \rscite{Dyhd16DME,Nava18DMEEDF} and use the (simplified) phase-space averaging (PSA) DME~\cite{Gebr10DME,Gebr11DME}.
The DME is applied to the isoscalar and isovector parts of the one-body density matrix using an isoscalar momentum scale, which works well for the former, but not for the latter~\cite{Zure21DMEs}.
However, the isovector Fock contributions are small and again we expect the Skyrme parameter fitting to partly compensate the errors.
We leave the investigation of the impact of choosing a different DME variant in the EDF construction for future work;
see \rcite{Zure21DMEs} for a study where similar tests are performed in a non-self-consistent scenario.
In that work we 
found DMEs work well even for pion exchanges at leading order (LO) in the chiral expansion despite the long range of this interaction.
Interaction terms at higher orders are of shorter range and therefore expected to be even more suited for a DME treatment.

Note that some of the 3N Fock terms used in \rcite{Nava18DMEEDF} were incorrect; these have been corrected in the present work.
We provide the resulting interpolation parameters entering \cref{eq:NNarctan,eq:3Narctan} 
in the Supplemental Material~\cite{GUDESupplemental}
and introduce the notation
\begin{equation}
    W_t^{u v}(\rho_0) = C_t^{u v}(\rho_0) + T_t^{u v} + \tilde g_t^{u v}(\rho_0) + \tilde g^{\rho_0 u_t v_t}(\rho_0) \rho_0
\end{equation}
for the combination of Skyrme coefficient, Taylor-expanded Hartree contribution, as well as NN and 3N $g$ coefficient functions of the same kind.

\subsection{\label{sec:pairing}Pairing contribution}

Within the HFB framework, the pairing contribution to our EDFs is given in the mixed-pairing prescription~\cite{Doba02Pairing} as
\begin{align}\label{eq:pairing}
    \Epair = \frac{1}{4} \sum_{q=\text{n}, \text{p}} \int \! \dd\vR \,
    V_0^q \left[ 1 - \frac{1}{2} \frac{\rho_0(\vR)}{\rho_s} \right] \tilde \rho_q^2(\vR) \,,
\end{align}
where $\tilde \rho_q(\vR)$ are the pairing densities and $\rho_s = 0.16 \unit{fm$^{-3}$}$.
The neutron and proton pairing strengths $V_0^\text{n}$ and $V_0^\text{p}$ are adjusted to data as described in \cref{sec:optimization}.
Because of the zero range of the underlying effective pairing force, a cutoff of $E_\text{cut} = 60\unit{MeV}$ to truncate the quasiparticle space is employed.
This cutoff was missing
in the implementation of \rcite{Nava18DMEEDF}.
Thus, in that work the quasiparticle space was truncated implicitly only, via the finite size of the employed basis.

In Ref.~\cite{Nava18DMEEDF}, we approximated particle number projection with a variant of the Lipkin-Nogami (LN) prescription derived for a seniority-pairing interaction with an adjusted effective strength~\cite{Stoi03drip}.
In \rcite{Stoi07SkyVAPNP} it was shown that this scheme compared well against the numerically expensive variation-after-projection scheme in well-deformed nuclei, but not near closed shells; see also \rcite{Bert09OddEven}. 
In addition to the lack of consistency between the actual pairing interaction and the one used for the LN scheme, the LN scheme is not variational. 
For these reasons, 
we drop this prescription and work at the HFB level only.
Future development of this work's EDFs should involve revisiting particle-number restoration.
Note that the UNEDF1-HFB parametrization of the Skyrme EDF was also performed without the seniority-based LN scheme of its parent UNEDF1 and its performance was only slightly worse~\cite{Schu15EDFError}.

\subsection{\label{sec:HFB}Hartree-Fock-Bogoliubov calculations}

We obtain nuclear ground states based on the EDFs described in the previous subsections by performing HFB calculations.
The HFB equations are solved with the DFT code \texttt{HFBTHO}, which expands the single-particle wave functions in a harmonic-oscillator (HO) basis in cylindrical coordinates~\cite{Mare22HFBTHOv4}.
For calculations of ground states, bases without axial deformation are used.
In all cases the basis consists of 20 HO shells
and
the spherical frequency $\omega_0$ of the HO basis is set according to the empirical formula $\omega_0 = 1.2 \times 41/A^{1/3} \unit{MeV}$~\cite{Stoi13HFBTHOv2} unless noted otherwise.
HFB solutions are obtained iteratively using the kickoff mode of \texttt{HFBTHO} in which an axial quadrupole deformation constraint is applied during at most the first ten HFB iteration steps to guide the solution towards the correct local minimum, then the constraint is lifted~\cite{Stoi13HFBTHOv2,Nava17HFBTHOv3}.

\subsection{\label{sec:optimization}Optimization of Skyrme and pairing parameters}

$\ESk$ and $\Epair$ contain in total 15 parameters $C_t^{uv}$, $\gamma$, and $V_0^q$ which need to be determined from fitting to data.
Note that $\EH$ and $\EF$ are free of adjustable parameters.
Thus, the number of optimization parameters is the same for functionals constructed here with and without chiral terms.
The volume parameters $C_{t0}^{\rho\rho}, C_{tD}^{\rho\rho}, C_{t}^{\rho\tau}$, and $\gamma$ can be related to properties of infinite nuclear matter (INM).
Expressing the exponent $\gamma$ in terms of INM parameters at saturation gives
\begin{align}\label{eq:gamma}
    \gamma = &\left\{
   -\left(\Incomp - \Incomp_\text{fr}\right)
   -9 \left(\SatEn - E_\text{sat,fr} + \frac{P_\text{fr}}{\SatDens} \right) 
   \right. \notag \\
   &\quad\left.+ \frac{\hbar^2}{2m} \left[ 4 \left(\ScEffMassInv - M_\text{s,fr}^{\ast -1} \right) -3 \right] C \SatDens^{2/3} + A_{ \gamma }(u_c)
   \right\} \notag \\
   &\bigg/ \left\{
   9 \left(\SatEn - E_\text{sat,fr} + \frac{P_\text{fr}}{\SatDens} \right) \right. \notag \\
   &\quad\left. +\frac{3\hbar^2}{2m} \left[ 2 \left(\ScEffMassInv - M_\text{s,fr}^{\ast -1} \right) -3 \right] C \SatDens^{2/3} +B_{ \gamma }(u_c)
   \right\}
   \,, 
\end{align}
where quantities indexed ``$\text{fr}$'' represent the contributions from the finite-range Hartree terms to the INM parameters (see \rcite{Sell14GognyNM}).
$P$ denotes the pressure of symmetric matter at saturation density, $C = (3/5) (3 \pi^2 / 2)^{2/3}$, and $u_c=(3 \pi^2 \SatDens / 2)^{1/3} / m_\pi$.
The expressions for $A_{ \gamma }(u_c)$ and $B_{ \gamma }(u_c)$ are given in Appendix C of \rcite{Stoi10DMEEDF}.
The equations for the other volume parameters can easily be obtained from the ones given in \rcite{Stoi10DMEEDF} by adding the respective contributions from the finite-range Hartree terms~\cite{Sell14GognyNM}.

Proceeding in this way we express the volume parameters via saturation density $\SatDens$, saturation energy $\SatEn$, incompressibility of symmetric nuclear matter $\Incomp$, isoscalar effective mass $M_\text{s}^\ast$, symmetry energy at saturation density $\SymEn$, its slope $\SlopePar$, and isovector effective mass $M_\text{v}^\ast$.
As in previous works~\cite{Kort10edf, Kort12UNEDF1, Kort14UNEDF2, Nava18DMEEDF} we do not optimize the isovector effective mass but instead keep it fixed at its SLy4 value, $\VecEffMassInv = 1.249$, which leaves 14 parameters to be optimized.

Using INM properties at saturation density as optimization parameters instead of EDF volume parameters allows us to impose physically motivated constraints on these parameters.
The bounds that we impose are not allowed to be violated in our optimization procedure.
We take the same bounds as in \rscite{Kort10edf, Kort12UNEDF1, Kort14UNEDF2, Nava18DMEEDF} except for $\Incomp$ and $\SlopePar$.
For the incompressibility $\Incomp$ we extend the allowed range to [180, 260] MeV based on the analysis of \rcite{Dris16asym} using different forces from chiral EFT, which obtained a range of [182, 262] MeV, and a study assessing the nuclear matter properties of Skyrme EDFs, which used [200, 260] MeV based on different experimental and empirical results~\cite{Dutr12SkyrmeNM}. 
For the slope parameter $\SlopePar$ we use [30, 80] MeV based on the overlapping region of different experimental and theoretical constraints; see \rscite{Dris20EOSBayes, Huth21ConstrEOS}.
Collectively we denote our set of optimization parameters as $\vx$.
The parameters and their ranges are summarized in \cref{tab:Bounds}.

\begin{table}[tbp]
    \caption{\label{tab:Bounds}
    Parameters optimized in this work and their bound constraints. 
}
\begin{ruledtabular}
\begin{tabular}{l|cc}
$\vx$ & Lower bound & Upper bound \\ \midrule 
$\SatDens \unit{(fm$^{-3}$)}$ & 0.15 & 0.17 \\ 
$\SatEn \unit{(MeV)}$ & $-16.2$ & $-15.8$ \\ 
$\Incomp \unit{(MeV)}$ & 180 & 260 \\ 
$\ScEffMassInv$ & 0.9 & 1.5 \\  
$\SymEn \unit{(MeV)}$ & 28 & 36 \\ 
$\SlopePar \unit{(MeV)}$ & 30 & 80 \\ 
$C_0^{\rho \Delta\rho}$ & $-\infty$ & $\infty$ \\
$C_1^{\rho \Delta\rho}$ & $-\infty$ & $\infty$ \\
$C_0^{\rho \nabla J}$ & $-\infty$ & $\infty$ \\
$C_0^{\rho \nabla J}$ & $-\infty$ & $\infty$ \\
$C_0^{JJ}$ & $-\infty$ & $\infty$ \\
$C_1^{JJ}$ & $-\infty$ & $\infty$ \\
$V_0^\text{n}$ & $-\infty$ & $\infty$ \\
$V_0^\text{p}$ & $-\infty$ & $\infty$ \\
\end{tabular} 
\end{ruledtabular}
\end{table}

They are determined by minimizing a \PenaltyFunc, which is given by a weighted sum of squared errors:
\begin{equation}
    \chi^2 (\vx) = \sum_{i=1}^{D_T} \sum_{j=1}^{n_i} \left( \frac{s_{i,j}(\vx) - d_{i,j}}{w_i} \right)^2 \,,
\label{eq:loss}
\end{equation}
where $s_{i,j}(\vx)$ are the EDF predictions and $d_{i,j}$ the data. 
$D_T$ is the number of different data types.
In this work we consider ground-state energies of spherical ($E_\text{sph}$) and deformed ($E_\text{def}$) nuclei, neutron ($\Delta_\text{n}$) and proton ($\Delta_\text{p}$) odd-even staggerings, proton point radii ($R_\text{p}$), and fission isomer excitation energies ($E^\ast$),
therefore $D_T = 6$.
For every data type $i$ we employ a different inverse weight $w_i$ that represents the expected errors in describing the different observables~\cite{Schu15EDFError}.
Rather than the somewhat arbitrary values set in \rcite{Nava18DMEEDF}, we choose for the weights the estimates determined from the Bayesian calibration of the UNEDF1 functional \cite{Schu20EDFCalibr}; see \cref{tab:Weights} for the numerical values.
This choice is justified by the fact that the data types contained in our fit data set are the same as for UNEDF1.
In addition, the form of the functionals (at least for our reference EDF without contributions from chiral EFT) as well as the employed optimization protocol are similar.

\begin{table}[bp]
\caption{\label{tab:Weights}
Characteristics of the components of the \PenaltyFunc. 
$n_i$  is the number of data points for each data type $i$ and $w_i$ is the inverse weight. For the latter, all units are MeV except $R_\text{p}$ which is in fm. 
}
\begin{ruledtabular}
\begin{tabular}{l|cccccc}
$i$ & $E_\text{sph}$ & $E_\text{def}$ & $\Delta_\text{n}$ & $\Delta_\text{p}$ & $R_\text{p}$ & $E^\ast$ \\ \midrule
$n_i$ & 29 & 47 & 7 & 6 & 28 & 4 \\
$w_i$ & 1.95 & 0.227 & 0.0457 & 0.0703 & 0.0177 & 0.85 \\
\end{tabular}
\end{ruledtabular}
\end{table}

\begin{figure}[tbp]
\includegraphics[width=1\linewidth]{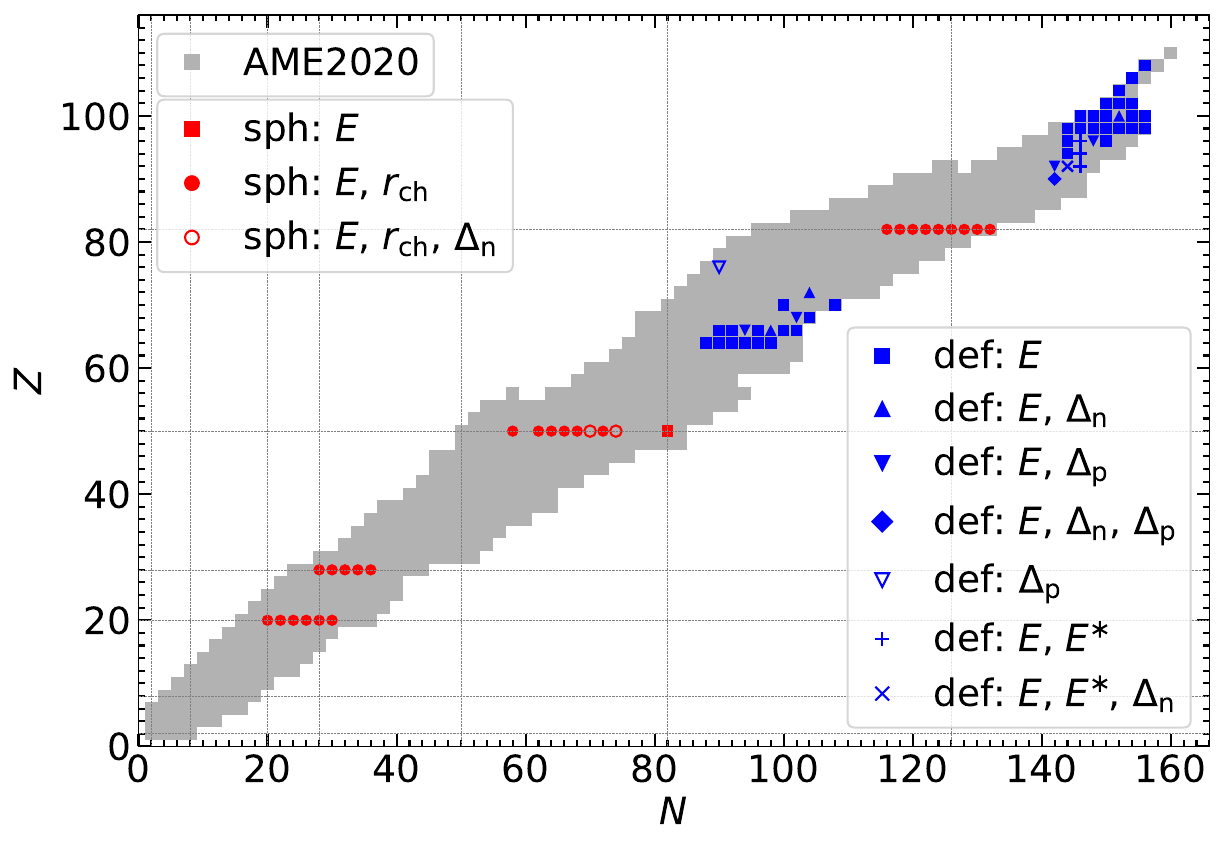}
\caption{\label{fig:fit_nuclei}
Experimental data used for optimization of EDF parameters. 
All even-even nuclei for which the ground-state binding energies are given in the 2020 Atomic Mass Evaluation~\cite{Wang21AME20} (excluding evaluated masses) are depicted in gray.
Nuclei included in the fit protocol are shown with different red and blue symbols depending on the considered data types. 
}
\end{figure}

\Cref{fig:fit_nuclei} shows in detail which data types are considered for which nuclei.
The experimental data is similar to the data used in \rscite{Kort14UNEDF2, Nava18DMEEDF}.
However, we exclude single-particle level splittings from the data set.
These were introduced in \rcite{Kort14UNEDF2} together with removing the restriction of $C_0^{JJ} = C_1^{JJ} = 0$ for the tensor part of UNEDF1 in an attempt to improve the description of nuclear shell structure.
The reported standard deviations for the tensor coefficients were quite large and the observed improvement of the shell structure relatively small.
Because the blocking calculations carried out to determine the single-particle structure are numerically expensive, we therefore decide to remove the single-particle level splittings from the data set.

With those exceptions, we consider the same data types for the same nuclei as in \rscite{Kort14UNEDF2, Nava18DMEEDF}. 
The experimental binding energies -- which determine $E_\text{sph}$, $E_\text{def}$, $\Delta_\text{n}$, and $\Delta_\text{p}$ -- are extracted from the 2020 Atomic Mass Evaluation (AME)~\cite{Wang21AME20} and the charge radii from \rcite{Ange13rch}; see \rscite{Kort14UNEDF2, Nava18DMEEDF} for details.
For \elm{Ni}{56}, which had not been measured yet, we take the value determined in \rcite{Somm22NiRch}. 
The conversion from charge radius to proton point radius is based on the 2018 CODATA recommended value for the proton charge radius $r_\text{p} = 0.8414 \unit{fm}$~\cite{Ties21CODATA2018} and the 2022 Particle Data Group average for the neutron charge radius square $r_\text{n}^2 = -0.1155 \unit{fm}^2$~\cite{Work22PDGReview}. 
The fission isomer energies are taken from \rcite{Sing02FissIsom}.

The EDF predictions $s_{i,j}(\vx)$ are obtained for given values of the parameters $\vx$ at every optimization step by solving HFB equations with the setup explained in \cref{sec:HFB}.
The value of the quadrupole moment used to initialize the kickoff mode is computed by assuming a ground-state deformation of $\beta_2 = 0.3$ for deformed nuclei and a fission isomer deformation of $\beta_2 = 0.6$~\cite{Schu20EDFCalibr}.
In total, 81 HFB calculations are performed at every optimization step: 
77 for the ground states of the nuclei in the data set, for which no axial basis deformation is used, and 4 for the fission isomers, which are calculated with an axial basis deformation parameter of $\beta = 0.4$.

We use the predicted average neutron (proton) HFB pairing gap as a proxy for neutron (proton) odd-even staggering. 
While this is an approximation~\cite{Doba95HFBShell}, actually determining odd-even mass differences would require calculating ground states of odd nuclei for which additional EDF terms enter due to broken time-reversal invariance and the determination of odd ground states via blocking calculations is much more involved than calculating ground states of even-even nuclei~\cite{Schu10Blocking}.

\begin{table*}[tbp]
    \caption{\label{tab:Optima}Parameters of the different \gude variants obtained in this work. 
    Values that are underlined correspond to cases where the minimum was attained at a parameter bound.
    $\SatDens$ is given in fm$^{-3}$, $\SatEn$, $\Incomp$, $\SymEn$, and $\SlopePar$ are in MeV, the surface coefficients $C_t^{\rho\Delta\rho}$, $C_t^{\rho\nabla J}$, and $C_t^{JJ}$ are in MeV~fm$^5$, and the pairing strengths $V_0^\text{q}$  are in MeV~fm$^3$.
    The last row gives the value of the \PenaltyFunc \eqref{eq:loss} at the minimum. 
    }
\begin{ruledtabular}
\begin{tabular}{l|r|rr|rrrrrrr}
    & \multicolumn{1}{c|}{\classZero} &  \multicolumn{2}{c|}{\classOne} & \multicolumn{7}{c}{\classTwo}		\\
	&	\multicolumn{1}{c|}{	\nochi	}	&	\multicolumn{1}{c}{	LO	}	&	\multicolumn{1}{c|}{	NLO	}	&	\multicolumn{1}{c}{	\NNLO	}	&	\multicolumn{1}{c}{	\NNLOThreeN	}	&	\multicolumn{1}{c}{	\NLOD	}	&	\multicolumn{1}{c}{	\NLODThreeN	}	&	\multicolumn{1}{c}{	\NNLOD	}	&	\multicolumn{1}{c}{	\NNLODThreeN	}	&	\multicolumn{1}{c}{	\minchi	}	\\ \midrule
$\SatDens$	& $		0.15463		$ & $		0.15430		$ & $		0.15423		$ & $		0.15779		$ & $		0.15749		$ & $		0.15571		$ & $		0.15615		$ & $		0.15606		$ & $		0.15681		$ & $		0.15832		$ \\
$\SatEn$ 	& $	\underline{	-15.8	}	$ & $	\underline{	-15.8	}	$ & $	\underline{	-15.8	}	$ & $	\underline{	-15.8	}	$ & $	\underline{	-15.8	}	$ & $	\underline{	-15.8	}	$ & $	\underline{	-15.8	}	$ & $	\underline{	-15.8	}	$ & $	\underline{	-15.8	}	$ & $		-15.830		$ \\
$\Incomp$	& $	\underline{	260	}	$ & $	\underline{	260	}	$ & $	\underline{	260	}	$ & $		222.2		$ & $		215.2		$ & $		240.8		$ & $		230.9		$ & $		236.0		$ & $		222.4		$ & $		223.6		$ \\
$\ScEffMassInv$	& $		0.9788		$ & $		0.9579		$ & $		0.9641		$ & $		0.9048		$ & $		0.9027		$ & $	\underline{	0.9	}	$ & $	\underline{	0.9	}	$ & $	\underline{	0.9	}	$ & $		0.9057		$ & $		0.9173		$ \\
$\SymEn$	& $		29.95		$ & $		30.98		$ & $		30.99		$ & $		28.07		$ & $		28.45		$ & $		28.43		$ & $		28.63		$ & $		28.37		$ & $		28.60		$ & $		28.58		$ \\
$\SlopePar$	& $		41.4		$ & $		59.6		$ & $		58.9		$ & $		34.1		$ & $	\underline{	30	}	$ & $	\underline{	30	}	$ & $	\underline{	30	}	$ & $	\underline{	30	}	$ & $	\underline{	30	}	$ & $	\underline{	30	}	$ \\
$C_0^{\rho\Delta\rho}$	& $		-41.4		$ & $		-37.5		$ & $		-38.4		$ & $		24.5		$ & $		9.4		$ & $		18.5		$ & $		8.2		$ & $		27.0		$ & $		10.9		$ & $		22.5		$ \\
$C_1^{\rho\Delta\rho}$	& $		-6.4		$ & $		-25.0		$ & $		-15.1		$ & $		-83.2		$ & $		-21.6		$ & $		-12.9		$ & $		-3.4		$ & $		-17.3		$ & $		-5.6		$ & $		-38.8		$ \\
$C_0^{\rho\nabla J}$	& $		-62.3		$ & $		-72.9		$ & $		-74.2		$ & $		-82.6		$ & $		-88.3		$ & $		-65.5		$ & $		-77.7		$ & $		-65.3		$ & $		-86.3		$ & $		-61.4		$ \\
$C_1^{\rho\nabla J}$	& $		11.0		$ & $		18.1		$ & $		15.5		$ & $		-39.3		$ & $		18.6		$ & $		17.5		$ & $		23.5		$ & $		14.9		$ & $		19.7		$ & $		3.4		$ \\
$C_0^{JJ}$	& $		-43.4		$ & $		-75.1		$ & $		-75.8		$ & $		-53.4		$ & $		-78.1		$ & $		-100.4		$ & $		-97.4		$ & $		-103.3		$ & $		-83.7		$ & $		-38.8		$ \\
$C_1^{JJ}$	& $		-30.1		$ & $		-15.0		$ & $		-12.3		$ & $		12.3		$ & $		1.3		$ & $		-10.2		$ & $		-8.0		$ & $		-11.0		$ & $		-2.6		$ & $		-4.2		$ \\
$V_0^\text{n}$	& $		-218.4		$ & $		-219.9		$ & $		-220.9		$ & $		-207.2		$ & $		-209.1		$ & $		-205.8		$ & $		-207.2		$ & $		-206.5		$ & $		-209.1		$ & $		-206.5		$ \\
$V_0^\text{p}$	& $		-259.9		$ & $		-263.0		$ & $		-263.2		$ & $		-246.4		$ & $		-255.5		$ & $		-251.9		$ & $		-253.7		$ & $		-252.5		$ & $		-255.3		$ & $		-249.4		$ \\ \midrule
$\gamma$	& $		0.467		$ & $		0.546		$ & $		0.541		$ & $		0.358		$ & $		0.320		$ & $		0.432		$ & $		0.385		$ & $		0.418		$ & $		0.352		$ & $		0.363		$ \\ \midrule
$\chi^2$	& $		122.4		$ & $		144.9		$ & $		145.5		$ & $		89.3		$ & $		88.7		$ & $		86.2		$ & $		89.1		$ & $		86.5		$ & $		90.7		$ & $		87.4		$
\end{tabular}
\end{ruledtabular}
\end{table*}

To find the parameter set $\vx$ for which $\chi^2(\vx)$ is minimized within the bound constraints discussed above we employ the derivative-free optimization algorithm \texttt{POUNDERS}~\cite{Wild15POUNDEDF,Wild17POUNDERS}.
It solves the nonlinear least squares problem by constructing a quadratic model for each term in the $\chi^2$. 
The resulting quadratic model for the $\chi^2$ is assumed to be valid only within a certain trust region.
Minimizing the model in this region yields a solution candidate point.
Then new quadratic models are constructed around this point and the trust region is updated.
In this way an iterative optimization procedure is obtained; see \rcite{Wild17POUNDERS} for details on the algorithm.
\texttt{POUNDERS} needs significantly fewer iteration steps to converge to a minimum than a conventional Nelder-Mead optimization routine \cite{Kort10edf,Wild15POUNDEDF}.

At every iteration step, the trust region is essentially a hypersphere around the current candidate point (in a space where the different optimization parameters are scaled as described in \rcite{Wild15POUNDEDF}).
The hypersphere's radius shrinks when getting closer to the minimum.
Sometimes \texttt{POUNDERS} shrinks this radius too quickly despite the current candidate point not being sufficiently close to the optimum yet.
In such scenarios, restarting \texttt{POUNDERS} from the current candidate point helps to accelerate the convergence
and allows it to possibly jump to another valley in the parameter landscape.
Therefore, we restart the optimization 
every 150 iteration steps and in doing so set the trust region radius back to its initial value of $\Delta_0 = 0.1$.

We use the parameter sets obtained at different orders in the chiral expansion in \rcite{Nava18DMEEDF} as starting points for the optimization of the corresponding \gude functionals constructed here.
For the reference ``\nochi'' functional we start the optimization from the UNEDF2 parameters~\cite{Kort14UNEDF2}. 
For a few EDFs we carry out the optimizations more than once employing also other Skyrme parametrizations as starting points (e.g., SLy4~\cite{Chab98SLy}).
We find that if those optimization runs converge, they converge to the same solutions as the other optimizations.
This gives us confidence that the parametrizations we obtain constitute global optima (within the employed bound constraints).

\section{\label{sec:results}Results}

\subsection{\label{sec:parametrizations}\gude parametrizations}

The parameter values obtained from the optimizations described in \cref{sec:optimization} are given in \cref{tab:Optima}. 
Parameters that ended up at their bounds are underlined.
We provide the EDF parameters with larger precision in the Supplemental Material~\cite{GUDESupplemental}, both in their explicit representation and equivalently in terms of INM properties.
We refer to the Skyrme-type \gude functional without any chiral terms as ``\nochi''.
The other \gude EDFs are labeled according to up to which order chiral terms are included and whether they include interaction terms with explicitly resolved intermediate $\Delta$ excitations and 3N forces.
We categorize the EDFs according to their properties discussed in the next paragraphs:
we refer to the ``\nochi'' functional as \classZero, to the LO and next-to-leading order (NLO) functionals collectively as \classOne, and to the remaining functionals as \classTwo.
This latter class contains also a functional labeled ``\minchi''.
It is constructed with the idea of adding as few terms as possible to the ``\nochi'' version while still obtaining an EDF that behaves like a member of \classTwo.
Details of the construction of this functional are discussed in \cref{sec:minchi}.
In \cref{tab:Optima}, the different classes are indicated by vertical lines.

We start with a discussion of the INM parameters of the different \gude variants.
The saturation energy $\SatEn$ ends up at its upper bound\footnote{
Note that for the \NNLOThreeN (obtained value of $\SatEn$: $-15.801$~MeV) and \NNLOD ($-15.8001$~MeV) 
EDFs the value of $\SatEn$ did not quite end up at the bound when the optimizations finished. 
Given that these values are very close to the bounds we expect that letting \texttt{POUNDERS} run longer would lead to parameter sets where these parameters are right at the bound.
}
for almost all optimized functionals.
This also holds for the value of the incompressibility $\Incomp$ for \classesZeroOne.
For \classTwo the incompressibility acquires lower values inside the allowed parameter range.
All other considered nuclear matter parameters also indicate a qualitative difference between \classesZeroOne on the one hand and \classTwo on the other hand:
the variation of the INM parameters within these groups is much smaller than the difference between them.
The main parameter difference between \classZero and \classOne lies in an increased value of the slope parameter $\SlopePar$ for the chiral functionals.
When going to the \classTwo functionals, $\SlopePar$ gets significantly reduced and ends up at its lower bound for most of the EDFs, with a correspondingly lower $\SymEn$ parameter.
Note that for some of the EDFs the inverse isoscalar effective mass $\ScEffMassInv$ attains its lower bound, too.
While $\ScEffMassInv=0.9$ is relatively low compared to typical values~\cite{Dutr12SkyrmeNM}, this value was also obtained for UNEDF0~\cite{Kort10edf}.

\begin{figure}[!bp]
\includegraphics[width=1\linewidth]{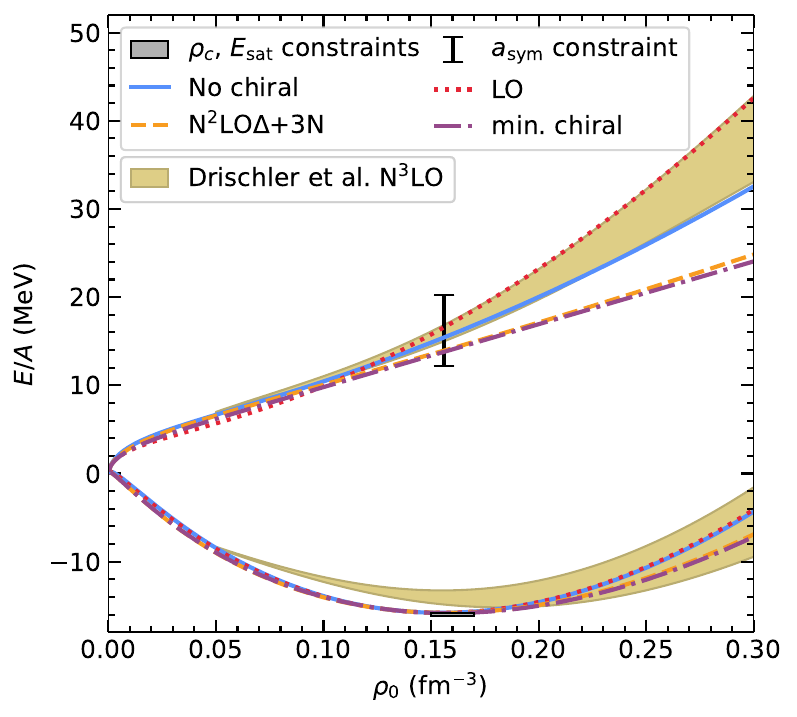}
\caption{\label{fig:infinite}
Energy per particle in infinite nuclear matter for selected \gude functionals constructed in this work.
For each EDF, both pure neutron matter and symmetric nuclear matter energies are shown.
The bound constraints on saturation density, saturation energy, and symmetry energy employed in the optimization of the EDFs are also depicted.
For comparison, we show the 1$\sigma$ uncertainty bands from a calculation employing a chiral Hamiltonian by Drischler et al.\ for $\rho_0 \geqslant 0.05 \unit{fm$^{-3}$}$~\cite{Dris20EOSBayes}.
}
\end{figure} 

In \cref{fig:infinite} we show the energy per particle for pure neutron matter and symmetric nuclear matter for four functionals constructed in this work; one each from \classZero and 1 and two from \classTwo.
The differences between the EDFs are very small up to about saturation density. 
This is not surprising since this region is probed by finite nuclei and hence strongly constrained by the fit to experimental data.
The differences between the different classes become much more pronounced for $\rho_0 \gtrsim \SatDens$, in particular for neutron matter. 
This region is not probed by finite nuclei, which is also why the deviation from the additionally given ab initio result observed for \classTwo in this density regime for neutron matter is not surprising. 
The plotted uncertainty bands have been obtained by Drischler et al.~\cite{Dris20EOSBayes} based on the MBPT calculations from Ref.~\cite{Dris19N3LO} with a chiral NN+3N Hamiltonian at \NNNLO with momentum cutoff 500~MeV~\cite{Ente17EMNn4lo} and 3N forces fit to saturation.
The difference of the EDFs from the ab initio results that is visible for symmetric matter is a consequence of the saturation energy bounds employed in this work, which are not obeyed by the ab initio results.
Note that the curves for the two \classTwo representatives, the \NNLODThreeN and the ``\minchi'' variant, are very close to each other even for $\rho_0 > \SatDens$.
This holds analogously for other EDFs from the same class.

Overall, and in particular within the classes as defined above, the description of INM at saturation density and below shows a large consistency between the different functionals.
This may be considered surprising given that the chiral contributions are quite different in size depending on the chiral order. 
However, it indicates that the optimization of the Skyrme and pairing coefficients to data can, to a large degree, wash out the effect of the additional terms.
We return to this issue in \cref{sec:analysis}.

In \cref{tab:Optima} we also provide the value of the $\gamma$ exponent for the different EDFs.
Compared to the ``\nochi'' variant it is larger for \classOne, but smaller for \classTwo, indicating that the density-dependent terms absorb different physics for the two classes.
Along the same lines we note that at every order $\gamma$ is smaller by about 0.05 for functionals including chiral 3N contributions.

For all \gude variants the generally observed hierarchy of pairing strengths $|V_0^\text{p}| > |V_0^\text{n}|$~\cite{Gori06BSk1013,Bert09OddEven} holds.
The somewhat weaker strengths obtained for the \classTwo EDFs when comparing to the other classes is in agreement with the lower $\ScEffMassInv$ values for \classTwo\cite{Bend03RMP}.

Note that a direct comparison of the surface parameters of the different \gude variants makes little sense because the chiral contributions to the corresponding terms depend on the functional and are not included in the $C_t^{u v}$ values given in \cref{tab:Optima}.

Based on starting optimization runs of the same \gude variant from different initial points~\cite{Schu15UQinDFT} we find that the parameters of the isovector part of the EDF
are relatively ill-constrained with our optimization protocol.
This is in agreement with observations made in other nuclear EDF optimizations~\cite{Kort10edf,Gao13SkyUncer,Kort14UNEDF2,Sell14GognyNM,McDo15UncQuaEDF}.
To better determine the isovector parameters the optimization data set has to be augmented; see also \cref{sec:SummaryOutlook}.
Also the $C_0^{JJ}$ parameter seems poorly constrained.
To quantify these statements a rigorous statistical analysis should be carried out in future work.

The last row of \cref{tab:Optima} contains the value of the \PenaltyFunc $\chi^2$ at the optimum.
For the ``\nochi'' EDF it is around 120.
Adding the chiral terms at LO (and NLO) according to the construction described in \cref{sec:method} worsens the $\chi^2$ at the minimum: it attains values around 145.
This stems from a slightly worse description of ground-state and fission isomer energies.

However, the additional inclusion of chiral terms at \NNLO or of the $\Delta$ contributions at NLO reduces the $\chi^2$ at the minimum to about 90.
In particular experimental energies of spherical nuclei in the fitting set are better described by the \classTwo functionals.
The root-mean-square deviation (RMSD) for those is 2.5~MeV for the ``\nochi'' EDF, but only 1.6~MeV for the \classTwo \gude variants.
The other data types in the $\chi^2$ are typically either slightly improved or are equally well described when comparing to the ``\nochi'' functional.

We note that the \NNLO EDF constitutes a slight deviation to these general trends (which can also been seen from some of the parameter values listed in \cref{tab:Optima}):
it describes the radii in the $\chi^2$ worse than all other EDFs but proton odd-even staggerings are much improved.

\subsection{\label{sec:minchi}Investigation of \gude \classTwo and construction of ``\minchi'' functional}

As discussed in the previous section and further in \cref{sec:global} we observe an improvement over the ``\nochi'' functional when going to EDFs that include chiral terms entering at \NNLO (or NLO when including interactions with explicit $\Delta$ excitations).
It turns out that only a small subset of the terms that contribute at these orders is actually necessary to achieve the improvement.

First, the inclusion of chiral isovector contributions is not required.
This is hardly surprising given that the Skyrme part of the EDFs contains six parameters contributing solely to the isovector part, which is to be compared to seven parameters for the isoscalar terms, but the isoscalar energy contributions are at least an order of magnitude larger than the isovector ones~\cite{Carl10DMEConv}.
The similar number of parameters for the two EDF parts suggests one may expect a similar relative precision for the corresponding energy contributions.
The resulting absolute deviations would then be much bigger for isoscalar energies.
Thus, one can expect omitting chiral isovector contributions does not significantly impact the description of bulk properties of finite nuclei (after refitting the EDF parameters).
Of course this is amplified by the inadequacy of the optimization data set to accurately fix the EDF isovector parameters. 

\begin{table}[tbp]
    \caption{\label{tab:HartreeDiff} 
    Exact scalar Hartree energies and differences of scalar Hartree energies calculated with Taylor expansions of the densities up to a given order 
    [cf. \cref{eq:HartreeTaylor}]
    and the corresponding exact energies (all in MeV). 
    The densities are generated from calculations with the SLy4 EDF.
    Results are given for the chiral pion exchanges considered here at \NNLO and for the finite-range parts of the Gogny D1S functional~\cite{Berg91GognyD1S}.}
\begin{ruledtabular}
\begin{tabular}{lc|c|ccc}
    & & & \multicolumn{3}{c}{Differences at order} \\
	  Interaction & Nucleus & Exact energy & 0 & 2 & 4 \\
	\midrule
  		\multirow{2}{*}{Chiral \NNLO} & \elm{Ca}{48} & $-759$ & $-118$ & $22$ & $-9$ \\
	& \elm{Pb}{208} & $-3937$ & $-290$ & $40$ & $-15$ \\
    \midrule
    \multirow{2}{*}{Gogny D1S} & \elm{Ca}{48} & $-9827$ & $-433$ & $27$ & $-4$ \\
    & \elm{Pb}{208} & $-47695$ & $-1028$ & $49$ & $-7$
\end{tabular}
\end{ruledtabular}
\end{table}

Performing an optimization of an EDF as described by \cref{eq:FullEDF} but taking into account from the chiral side only Fock contributions up to \NNLO yields a class-1-like functional which suggests that the switch to \classTwo is due to the Hartree terms.
Indeed \NNLO (NLO with $\Delta$s) is the first order which for even-even systems has isoscalar pion-exchange Hartree contributions.
These are by far the largest chiral contributions to the energy.
In \cref{tab:HartreeDiff} we show the expectation values of the exact Hartree energy from pion-exchange contributions up to \NNLO in the chiral expansion.
They are obtained with densities generated from calculations performed with the code \texttt{HOSPHE}~\cite{Carl10hosphe} employing the SLy4 EDF~\cite{Chab98SLy}.
Additionally, we provide the difference to these exact values for energies that we obtain when Taylor expanding one density entering the Hartree energy; see \cref{eq:HartreeTaylor}.
For comparison we also provide the analogous numbers obtained with the finite-range parts of the Gogny EDF in the D1S parametrization; see \rcite{Carl10DMEConv} for a more extensive study. 

One can see that the energies obtained with the Taylor series converge relatively slowly towards the exact values.
In particular when going to second order in the Taylor expansion the approximated value is still off by about 40 MeV in \elm{Pb}{208}.
The second-order expression for the energy has a Skyrme-like structure (density bilinears consisting of up to second-order densities multiplied with constant prefactors).
Therefore, one may expect that a Skyrme EDF cannot fully account for the chiral Hartree contributions at \NNLO if they are left out (as is the case for \classesZeroOne).
It is thus conceivable that \classTwoHyphen \gude variants behave differently from \classesZeroOne.\footnote{
Note that the argument put forward above is not a direct proof because the fitting of the EDF parameters may shuffle around contributions among more terms than the ones technically entering the Taylor-expanded energy.}

Carrying out the optimization of an EDF where in the chiral part only the isoscalar Hartree contributions entering at \NNLO are included leads to a functional with $\chi^2 \approx 112$ at the minimum, which is clearly larger than the values observed for \classTwo.
For this EDF the pairing strengths take a nonphysical value $V_0^q \approx 40 \unit{MeV fm$^3$}$.
These observations suggest another term is additionally needed to reproduce the \classTwoHyphen behavior.

\begin{figure}[tbp]
\includegraphics[width=1\linewidth]{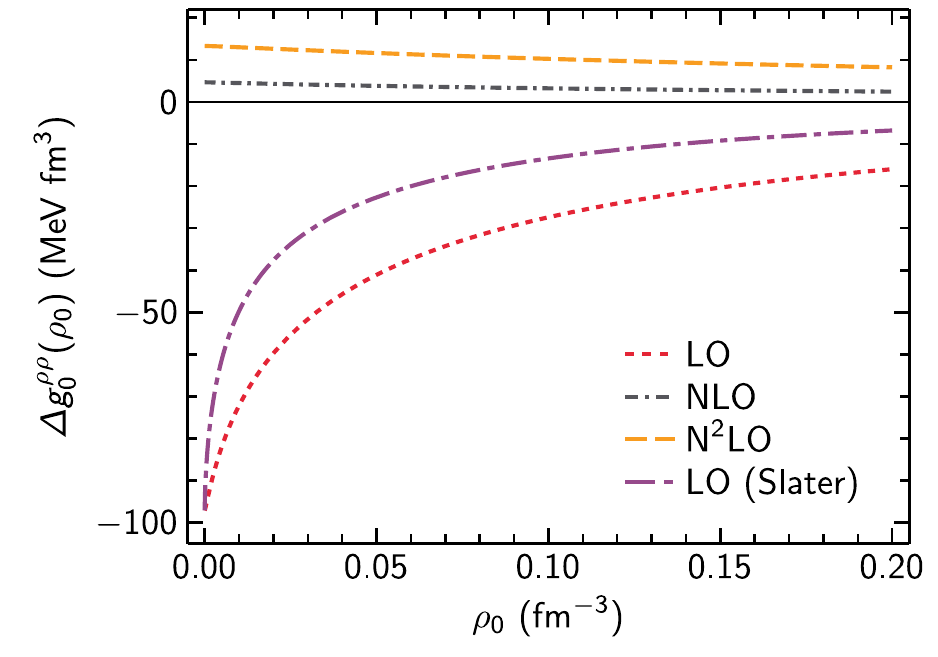}
\caption{\label{fig:rho_coeff}
Contributions to $g_0^{\rho\rho}$ arising at different chiral orders. 
We show contributions at LO, NLO, and \NNLO calculated from the interaction specified in \cref{sec:interaction} using the PSA-DME.
In addition, we show the LO contribution when using the Slater approximation instead of the PSA-DME.
}
\end{figure}

In \cref{fig:rho_coeff} we show the contributions to the $g_0^{\rho\rho}$ coefficient arising at different chiral orders,
but the following discussion applies also similarly to other $g$ coefficients.
The total $g_0^{\rho\rho}$ coefficient at a given order is the sum of all depicted contributions $\Delta g_0^{\rho\rho}$ up to that order.
The LO contribution shows a strong density dependence with its value at $\rho_0 = 0$ being about five times as large as the value at $\rho_0 = \SatDens$.
The contributions at NLO and \NNLO are much smaller and their density dependence is much weaker, which is why their effects can be easily captured by simply adjusting Skyrme coefficients.
In principle even the strongly density-dependent LO coefficient could be quite well mimicked by a Skyrme EDF due the presence of the $C_{tD}^{\rho\rho}\rho_0^\gamma$ term, but, since this term has to capture several different types of unresolved physics~\cite{Dugu03DensDep}, one may expect that adding the LO $g_0^{\rho\rho}$ contribution explicitly still has a relevant effect.
Optimizing an EDF with both isoscalar chiral long-range Hartree contributions at \NNLO and Fock contributions at LO yields a functional belonging to \classTwo as desired.

We showed in previous work~\cite{Zure21DMEs} that Fock energies from a Yukawa interaction can be well approximated by using the Slater approximation instead of the more involved PSA-DME applied in this work so far.
However, this comes at the price of a worse local reproduction of the Yukawa Fock energy density essentially everywhere in the nucleus.
Using the Slater approximation instead of the PSA-DME reduces the amount of nonvanishing isoscalar NN $g$ coefficient functions from five to one.
We show the nonzero $g_0^{\rho\rho}$ coefficient at LO in \cref{fig:rho_coeff}.
We find that the resulting EDFs differ by similar amounts as other functionals in \classTwo differ from each other.
Therefore it seems safe to use the simpler Slater approximation in the present EDF construction, at least for bulk properties.

We refer to the EDF constructed according to \cref{eq:FullEDF} including for $\EH$ only the isoscalar NN pion-exchange Hartree contribution entering at \NNLO and as $\EF$ the isoscalar NN pion-exchange Fock contribution at LO (described by the Slater approximation) as the ``\minchi'' \gude variant. 
The parameters obtained when optimizing this functional are given in \cref{tab:Optima} 
and with higher precision in the Supplemental Material~\cite{GUDESupplemental}, where we also provide
the parameters used in the interpolations for the chiral Hartree and Fock contributions according to \cref{eq:VCGaussians,eq:NNarctan}.
The INM parameters and the $\chi^2$ value at the optimum are in the ranges of the other \classTwoHyphen functionals (see \cref{tab:Optima}), indicating that the ``\minchi'' variant indeed also belongs to this class.
This explicitly demonstrates that the two identified terms are enough to achieve the improvement over \classesZeroOne.

\begin{table*}[tbp]
    \caption{\label{tab:rmsd}Deviations of ground-state energies, two-neutron and two-proton separation energies (all in MeV), and charge radii (in fm) calculated with the different \gude variants and the corresponding experimental values.
    The upper half of the table contains root-mean-square deviations, the lower half lists mean deviations.
    The values are calculated from all even-even nuclei with $Z \geqslant 8$ included in the experimental data sets,
    see text for details on those.
    }
\begin{ruledtabular}
\begin{tabular}{ll|r|rr|rrrrrrr}
  & & \multicolumn{1}{c|}{\classZero} &  \multicolumn{2}{c|}{\classOne} & \multicolumn{7}{c}{\classTwo}		\\
	 & &	\multicolumn{1}{c|}{	\nochi	}	&	\multicolumn{1}{c}{	LO	}	&	\multicolumn{1}{c|}{	NLO	}	&	\multicolumn{1}{c}{	\NNLO	}	&	\multicolumn{1}{c}{	\NNLOThreeN	}	&	\multicolumn{1}{c}{	\NLOD	}	&	\multicolumn{1}{c}{	\NLODThreeN	}	&	\multicolumn{1}{c}{	\NNLOD	}	&	\multicolumn{1}{c}{	\NNLODThreeN	}	&	\multicolumn{1}{c}{	\minchi	}	\\ \midrule
\multirow{4}{*}{RMSD}	&	$\GSEnergy$	& $		2.11		$ & $		2.09		$ & $		2.13		$ & $		1.56		$ & $		1.41		$ & $		1.47		$ & $		1.50		$ & $		1.42		$ & $		1.53		$ & $		1.45		$ \\
	&	$\NSepEnergy$	& $		0.86		$ & $		0.85		$ & $		0.89		$ & $		0.74		$ & $		0.73		$ & $		0.73		$ & $		0.75		$ & $		0.73		$ & $		0.77		$ & $		0.75		$ \\
	&	$\PSepEnergy$	& $		0.74		$ & $		0.74		$ & $		0.77		$ & $		0.61		$ & $		0.61		$ & $		0.60		$ & $		0.62		$ & $		0.59		$ & $		0.64		$ & $		0.63		$ \\
	&	$\ChargeRadius$ 	& $		0.024		$ & $		0.024		$ & $		0.025		$ & $		0.024		$ & $		0.023		$ & $		0.022		$ & $		0.022		$ & $		0.023		$ & $		0.023		$ & $		0.022		$ \\ \midrule
\multirow{4}{*}{Mean dev.}	&	$\GSEnergy$	& $		0.630		$ & $		0.532		$ & $		0.560		$ & $		0.324		$ & $		0.302		$ & $		0.369		$ & $		0.393		$ & $		0.296		$ & $		0.387		$ & $		0.293		$ \\
	&	$\NSepEnergy$	& $		0.089		$ & $		0.093		$ & $		0.094		$ & $		-0.060		$ & $		-0.015		$ & $		0.001		$ & $		0.010		$ & $		-0.006		$ & $		-0.003		$ & $		-0.015		$ \\
	&	$\PSepEnergy$	& $		-0.027		$ & $		-0.036		$ & $		-0.044		$ & $		0.153		$ & $		0.082		$ & $		0.064		$ & $		0.053		$ & $		0.073		$ & $		0.069		$ & $		0.081		$ \\
	&	$\ChargeRadius$ 	& $		-0.0092		$ & $		-0.0093		$ & $		-0.0090		$ & $		-0.0017		$ & $		0.0003		$ & $		-0.0018		$ & $		-0.0025		$ & $		-0.0006		$ & $		-0.0014		$ & $		-0.0010		$ \\
\end{tabular}
\end{ruledtabular}
\end{table*}

\subsection{\label{sec:global}Global comparison to experiment}
We now investigate the performance of the different functional variants in the \gude family obtained in \cref{sec:parametrizations} by calculating the ground states of even-even nuclei included in the 2020 AME~\cite{Wang21AME20}.
We include all 663 nuclei with actual measured masses, leaving out those for which only evaluated masses are available.
Every nucleus is calculated five times with \texttt{HFBTHO} in kickoff mode setting the initial deformation constraint to $\beta = -0.2, -0.1, 0, 0.1, 0.2$.
This is done so that oblate deformed, spherical, and prolate deformed solutions are considered as possible ground states for every nucleus.
The HFB calculations are carried out until they are converged (typically within at most about 100 HFB iteration steps) or until the amount of unconverged calculations for a given functional does not get further reduced for at least 800 HFB steps.
For most \gude variants only about a handful of the 
3315
calculations end up unconverged at the end of this procedure.
The \NNLO EDF is the only exception from this rule:
even after more than 3000 HFB steps, 111 calculations are still unconverged.
Note, however, that only four of those constitute the calculation with lowest binding energy for the corresponding nucleus.

For every nucleus, we pick among the converged calculations the one with the lowest energy as a first ground-state candidate and apply on it two filters to exclude unphysical solutions.
Whenever a filter is triggered, the calculation with the next-lowest energy for the same nucleus is considered instead.
First we do not consider solutions with $E/A < -11 \unit{MeV}$.
This filter turns out to be triggered only a few times by calculations with EDFs that include interactions with explicit $\Delta$ isobars in the chiral terms.
Second we apply a filter to remove solutions with unphysically large deformations.
This is done by applying the 1.5 interquartile range rule, which is a simple measure to detect outliers of a distribution,
on the values of the deformation parameter $\beta_2$ of all remaining ground state candidates.
The $\beta_2$ parameter is much less mass-number dependent than the axial quadrupole moment of the nucleus $Q_{20}$ and is related to it according to
\begin{equation}\label{eq:beta}
    \beta_2 = \sqrt{\frac{\pi}{5}} Q_{20} / \left(A R_\text{m}^2 \right) 
\end{equation}
with the root-mean-square matter point radius $R_\text{m}$. 
The deformation filter is in practice triggered at most for two nuclei per EDF.

We compare the resulting ground-state energies against the values extracted from the 2020 AME.
\Cref{tab:rmsd} contains the corresponding root-mean-square and mean deviations obtained for nuclei with $Z \geqslant 8$.
We also give the deviations of the two-neutron ($S_{2\text{n}}$) and two-proton ($S_{2\text{p}}$) separation energies obtained from the same data set, and of the charge radii from \rcite{Ange13rch}.
\gude variants of the same class behave very similar for all these quantities with the only exception being somewhat larger mean deviations observed for separation energies for the \NNLO functional compared to other \classTwoHyphen EDFs.

\begin{figure}[bp]
\includegraphics[width=1\linewidth]{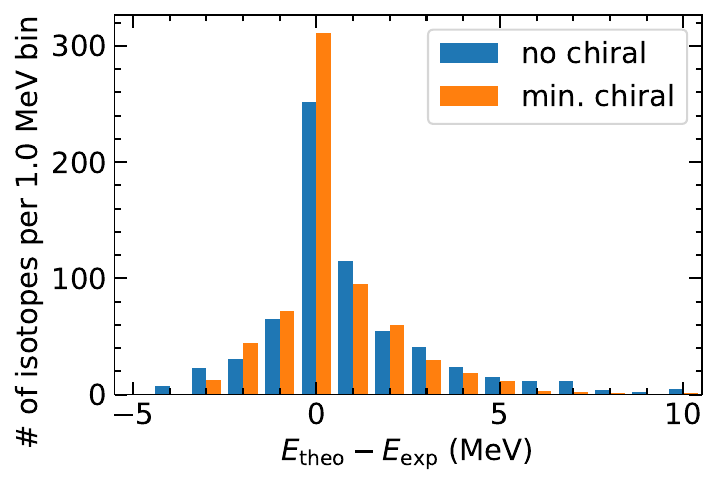}
\caption{Distributions of ground-state energy differences between calculated and experimental results.
They are shown for the ``\nochi'' and ``\minchi'' \gude functionals in bins with a width of 1~MeV each.
Note that the last bin contains also values with an energy difference larger than $10.5 \unit{MeV}$.
}
\label{fig:masses_histogram}
\end{figure} 

\begin{figure*}[tbp]
\includegraphics[width=1\linewidth]{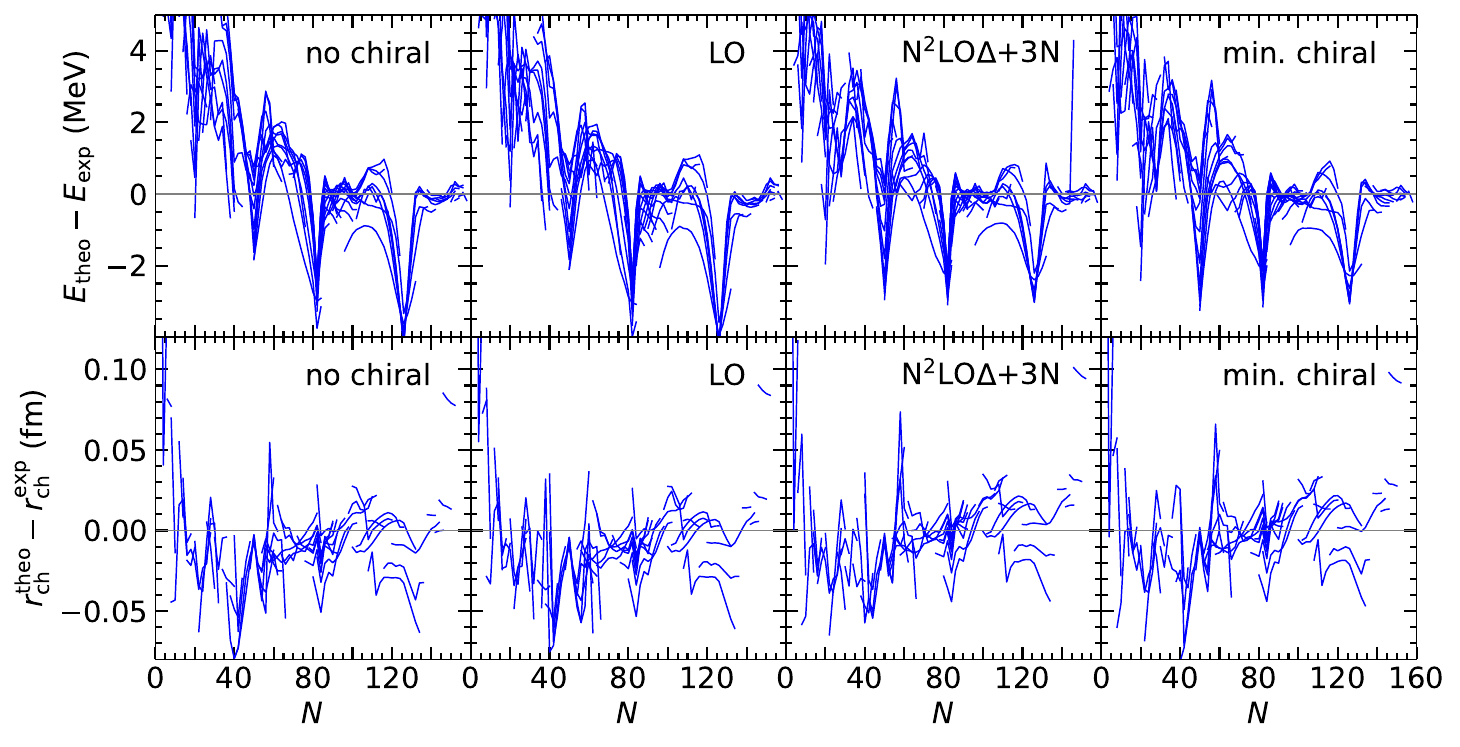}
\caption{\label{fig:global_masses}
Differences of ground-state energies (upper panels) and charge radii (lower panels) for even-even nuclei between values obtained with selected \gude variants and experiment.
See text for details on the experimental data.
}
\end{figure*} 

While \classesZeroOne perform similarly, an improvement is observed for all observables when going to \classTwo.
In particular, the ground-state energy RMSD is significantly reduced by roughly 30\% from 2.1~MeV for \classesZeroOne to about 1.5~MeV for the various \classTwoHyphen EDFs.
The mean deviation $\langle E_\text{theo} - E_\text{exp}\rangle$ is almost halved down to 0.3~MeV, indicating that the energies are less biased towards underbinding for \classTwo.
This can also be seen in \cref{fig:masses_histogram},%
\footnote{Note that this and following figures contain also nuclei with $Z<8$, unlike the values provided in \cref{{tab:rmsd}}.} 
which shows the histogram of the quantity $E_\text{theo} - E_\text{exp}$.
Calculations which produce extremely underbound nuclei (those at the very right of the distribution) occur much less often for the \classTwoHyphen ``\minchi'' functional than for the reference ``\nochi'' EDF.
Such cases correspond mostly to very light nuclei.
For the \classTwoHyphen variants almost half of all nuclei are described with a mass error of less than 0.5~MeV.
Note that while the binding energies included in the $\chi^2$ are described better by \classZero than by \classOne, the performance on all even-even nuclei binding energies is very similar for these two classes.

In the upper row of \cref{fig:global_masses} we show ground-state energy residuals for four \gude variants.
One can clearly see that the \classTwoHyphen EDFs describe energies around the $N=82$ and $N=126$ shell closures much better than the class-0 and -1 variants.
We note that, due to the parameter optimization involved in the construction of every functional, it is not clear if the additional chiral terms entering the \classTwoHyphen functionals are actually directly improving the description of (near-)closed-shell nuclei or if they instead improve the open shells and indirectly allow the parameter optimization to yield a better reproduction of closed shells.
In addition, the observed underbinding for light nuclei is reduced for the \classTwoHyphen variants.

For both two-neutron and two-proton separation energies, \classTwoHyphen EDFs give a small improvement over \classesZeroOne: 
the RMSD values are reduced by about 12\%.
In addition, the bias on $S_{2\text{n}}$ values is almost completely gone while it is increased for $S_{2\text{p}}$.

The description of charge radii is least affected by the additional chiral terms added in \classTwo.
This can also be seen in the lower row of panels of \cref{fig:global_masses}.
Charge radii are only slightly better described for $N \approx 40$ to $100$
and their mean deviation is slightly closer to zero for \classTwo.

\subsection{\label{sec:local}Shell structure and deformation properties
}

To investigate the quality of the \gude family with respect to nuclear shell structure, we compute single-particle levels using blocking calculations; see \rscite{Schu10Blocking,Kort12UNEDF1} for details on the procedure. 
Using blocking calculations at the HFB level is both logically consistent with the construction of the functionals at the HFB level and helps with reducing systematic errors when comparing with experiment~\cite{Kort12UNEDF1}.
Calculations use the same setting for the HO basis as before, namely with 20 full, spherical shells.
In this context one should be reminded that single-particle energies are not observables but extracted in a model-dependent way from experiment~\cite{duguet2012initio,duguet2015nonobservable}. 
Here we compare to the values given in \rcite{Schw07SPLevels}. 
Furthermore, it is well known that the single-particle shell structure depends strongly on beyond-mean-field effects such as particle-vibration couplings~\cite{colo2010effect,cao2014properties,tarpanov2014polarization,tarpanov2014spectroscopic}. 
As a consequence, blocking calculations should not be expected to perfectly match ``experimental'' single-particle data in closed shell nuclei. 
They are simply meant as a validation check to guarantee that basic features of the nuclear shell structure are properly reproduced.

\begin{figure}[tbp]
\includegraphics[width=1\linewidth]{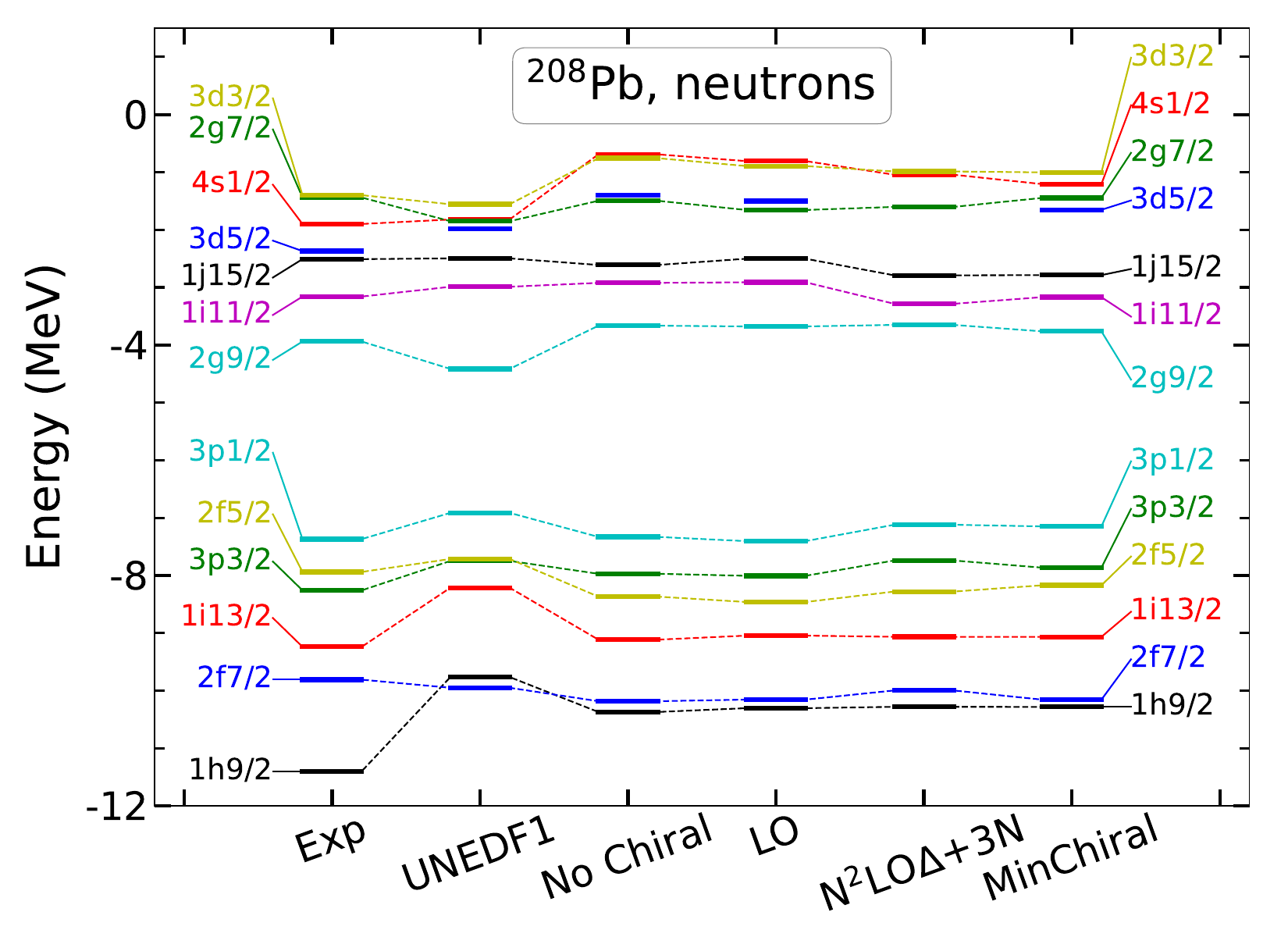}
\caption{\label{fig:lead_sp_n}
Single-particle spectrum for neutrons in $^{208}$Pb for a selection of EDFs.
}
\end{figure} 

As an illustrative example, we show in \cref{fig:lead_sp_n} the obtained neutron single-particle spectra of \elm{Pb}{208} for selected \gude EDFs representative of the different classes. 
One can make the following general observations. 
First, the single-particle levels turn out to be largely insensitive to the \gude variant. 
Second, the obtained shell gaps in \elm{Pb}{208} are in good agreement with the ones extracted from experiment and a little better reproduced than for the UNEDF1 functional.
Third, the level ordering of the occupied neutron orbitals is also in slightly better agreement with experiment.
These qualitative conclusions apply to other doubly closed shell nuclei and suggest a decent reproduction of the shell structure by the \gude functionals.

Next, we test deformation properties of the EDFs on the standard fission benchmark case of \elm{Pu}{240}.
The HFB calculations are carried out in a deformed  HO basis with $30$ shells included and with the HO frequency and basis deformation optimized for that nucleus; see \rcite{Schu14IndFiss} for details. 
A constraint on the octupole moment is imposed during the first ten iterations to ensure the fission goes through the most likely pathway. 
Calculations assume axial symmetry. 

\begin{figure}[tbp]
\includegraphics[width=1\linewidth]{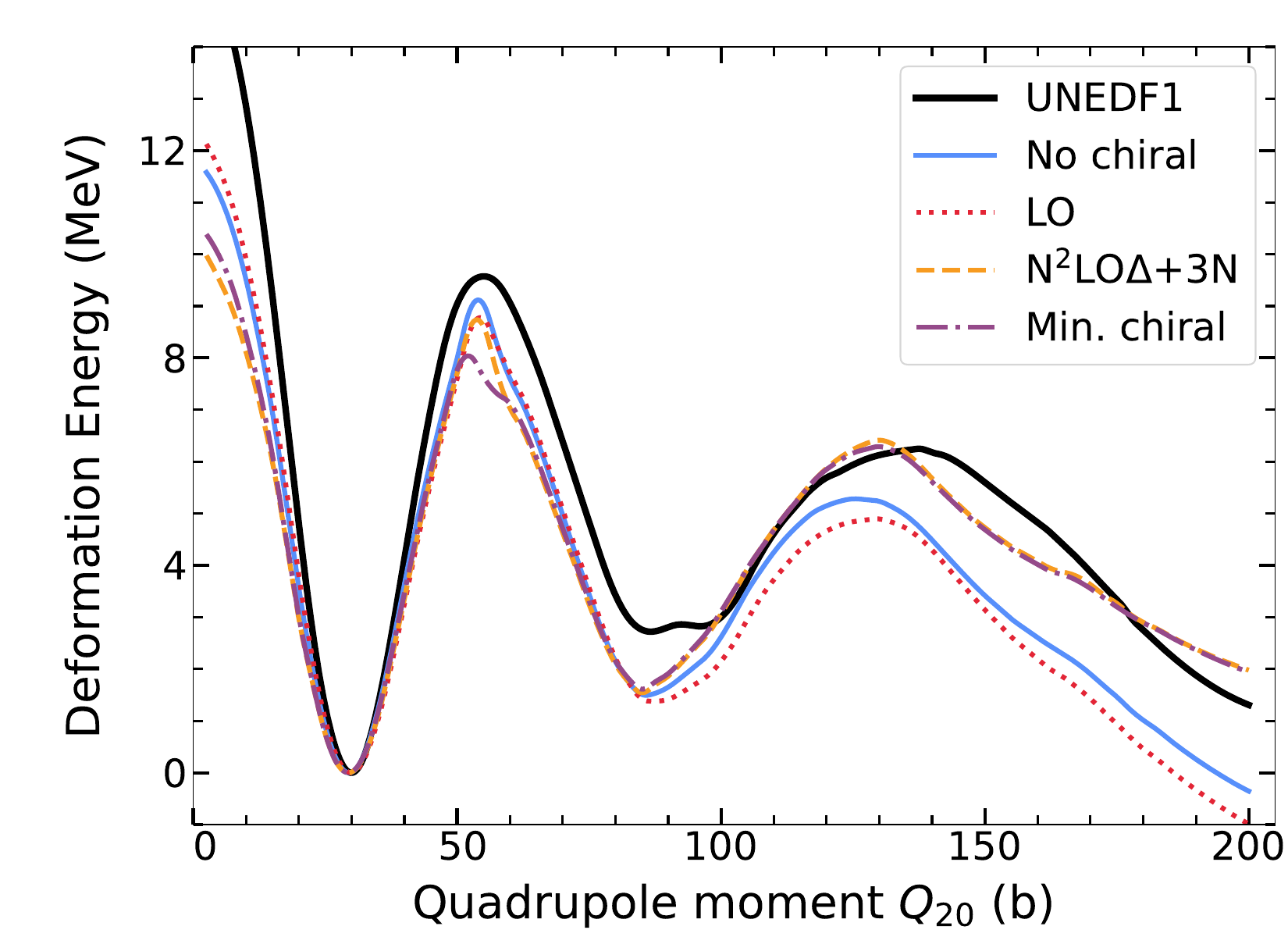}
\caption{\label{fig:deformation}
Deformation energy of $^{240}$Pu as a function of the axial quadrupole moment. Calculations assume axial symmetry.
}
\end{figure}

In \cref{fig:deformation} we show the deformation energy, i.e., the energy difference between the configuration with given deformation and the ground state, as a function of the quadrupole moment for selected \gude functionals as well as for UNEDF1 for comparison.
Since including triaxiality typically reduces the height of the first fission barrier by about 2~MeV~\cite{Giro83HFBTriax,Schu14IndFiss},
the overall agreement with values extracted from experiment~\cite{Capo09RIPL} is in fact very good for all considered \gude variants.
The energy of the fission isomer $E^\ast$ is predicted too low by about 1~MeV compared to the value used in the optimization set (2.8~MeV)~\cite{Sing02FissIsom}.
Seeing that the results for UNEDF1, UNEDF2, and the DME EDFs of \rcite{Nava18DMEEDF} agree very well with this experimental value, this is probably a consequence of the reduced weight of fission isomer energies in the present optimization protocol. 
Note that a newer experimental estimate for the fission isomer energy of $2.25\unit{MeV}$~\cite{Huny01Pu240r}
is closer to the \gude values. 

\begin{figure}[tbp]
\includegraphics[width=1\linewidth]{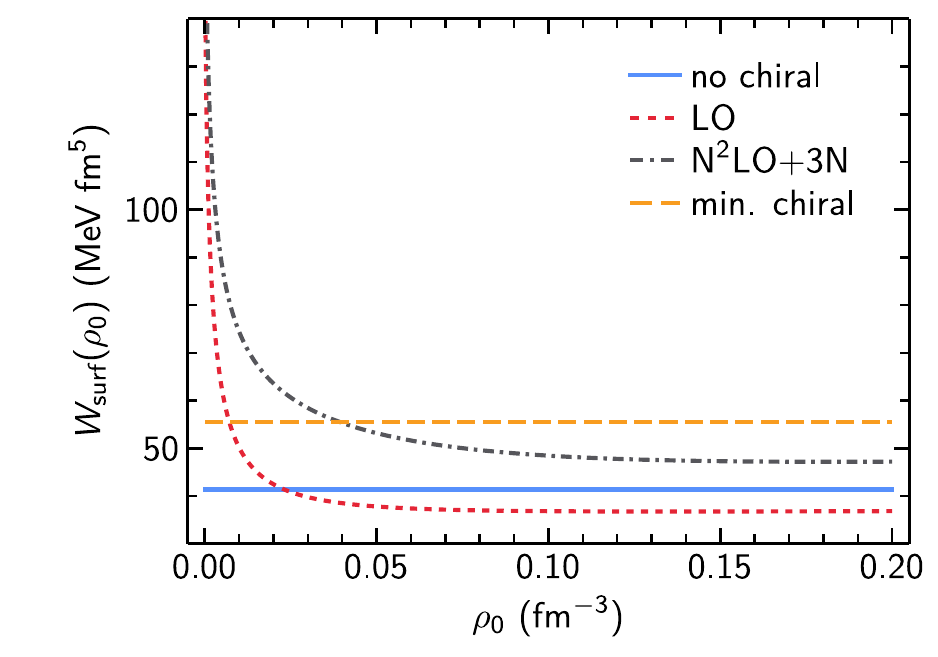}
\caption{\label{fig:W0rDr}
$W_\text{surf}$ for different \gude variants. 
}
\end{figure}

For values of $Q_{20}$ larger than the value at the fission isomer state a clear difference between results obtained for \classesZeroOne and \classTwo emerge as already observed for other quantities in this paper. 
We may speculate that such differences are the result of a competition between bulk and shell effects.
\Cref{tab:Optima} and \cref{fig:W0rDr} show that the symmetry energy $\SymEn$ and the surface coupling function 
$W_\text{surf}$ (defined below),
respectively, differ substantially for the class-0 and -1 and the \classTwoHyphen parametrizations.
For \classesZeroOne, the value of the symmetry energy is $\SymEn \approx 30$ MeV while it is $\SymEn \approx 28.5$ MeV for \classTwoHyphen EDFs.
The surface coupling function, which contains the full contribution to the isoscalar surface energy (Skyrme plus chiral terms), is given by
\begin{equation}
    W_\text{surf}(\rho_0) = W_{0}^{(\nabla\rho)^2}(\rho_0) + W_{0,\int}^{\rho\Delta\rho}(\rho_0) \,,
\end{equation}
where 
\begin{equation}
     W_{0,\int}^{\rho\Delta\rho}(\rho_0) = - W_{0}^{\rho\Delta\rho}(\rho_0) - \frac{\partial W_{0}^{\rho\Delta\rho}(\rho_0)}{\partial \rho_0} \rho_0 \,
\end{equation}
arises from integrating by parts:
\begin{equation}
    \int \! \dd\vR \, W_{0}^{\rho\Delta\rho}(\rho_0) \rho_0 \Delta \rho_0 = \int \! \dd\vR \, W_{0,\int}^{\rho\Delta\rho}(\rho_0) \vnabla\rho_0 \cdot \vnabla\rho_0 \,.
\end{equation}
$W_\text{surf}$ is for intermediate densities much stronger for \classTwoHyphen functionals than for \classesZeroOne.
Together, $\SymEn$ and $W_\text{surf}$ impact the surface and surface-symmetry contributions to the bulk energy, which are known to be key drivers of deformation properties~\cite{nikolov2011surface,jodon2016constraining}.
At the same time, \cref{fig:lead_sp_n} also shows a small but visible difference in the neutron shell structure between \classTwo and the other \gude variants functionals, with the $N=126$ shell gap being a little smaller for \classTwo. 
Such differences will be amplified as deformation increases and this could play a role in the deformation energy.

\section{\label{sec:analysis}Analysis of chiral contributions}

In this section we analyze why 
the only significant effects we obtain from including chiral interactions explicitly into the \gude functionals occur for the switch from \classOne to \classTwo, i.e., at \NNLO (NLO when including $\Delta$ isobars explicitly) in the chiral expansion.

\begin{figure}[tbp]
\includegraphics[width=1\linewidth]{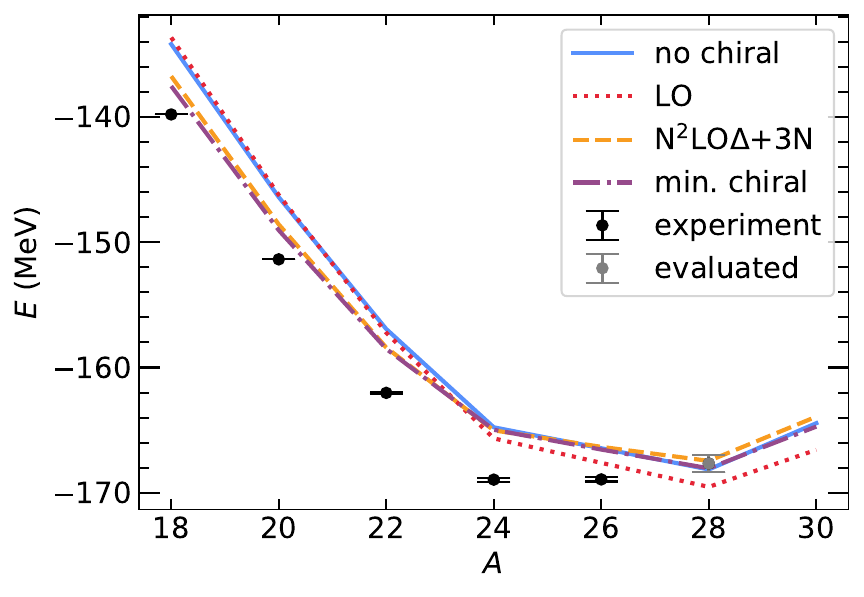}
\caption{\label{fig:oxygen_masses}
Ground-state energies of oxygen isotopes for selected \gude variants.
We also show experimental and evaluated results as provided in \rcite{Wang21AME20}.
}
\end{figure}

As stated in \cref{sec:results}, only little change over the reference ``\nochi'' EDF is seen when going to LO in the present construction; see especially \cref{tab:rmsd}.
This is not surprising since one-pion exchange is known to largely average out for bulk properties~\cite{Gelm95ChiralNM,Stoi10DMEEDF} because at this order pions enter at the mean-field level only through Fock contributions, which are small. 
For nonbulk quantities such as behaviors along isotopic chains, small differences between the ``\nochi'' and LO EDFs are visible; see for instance the oxygen chain shown in \cref{fig:oxygen_masses}.

At NLO pions enter at the HF level only through Fock and isovector Hartree contributions. 
Since these are very small and can be captured well by Skyrme terms due to the weak density dependence of the resulting $g$ coefficients (see, e.g., \cref{fig:rho_coeff}), the almost identical performance of the LO and NLO functionals is to be expected. 

When going to \NNLO a significant improvement is achieved, in particular for the global description of ground-state energies.
The detailed analysis of \cref{sec:minchi} indicates that the interplay of two contributions is responsible for this.
The attractive pion Hartree contribution at \NNLO is large and apparently cannot be completely mimicked by Skyrme terms only.
Its addition together with LO Fock terms leads to the improvement.

While the incompressibility is at its upper bound for \classesZeroOne, it is much smaller for the \NNLO EDF (and the other \classTwoHyphen ones); see \cref{tab:Optima}.
This is probably a consequence of the strongly attractive central isoscalar two-pion exchange entering at \NNLO in the chiral expansion~\cite{Hu17NucMatNN}.

This observation raises the question whether the additional chiral terms in \classTwo lead to a better description of experiment by themselves or whether the improvement is realized indirectly by moving the unbounded optimum ``closer'' to the bound constraint region and thereby reducing the achievable $\chi^2$ values within this region.
To address this issue one could perform an unconstrained optimization for the different \gude functionals.
Preliminary unconstrained optimizations suggest that the latter mechanism is the dominant one because the difference of the obtained $\chi^2$ values largely 
seems to vanish for the unbounded optima.
Note, however, that these conclusions are preliminary, since for some of the EDFs competing minima seem to occur during the unbounded optimizations
and sometimes the unconstrained optima seem to correspond to situations where some INM parameters attain values far away from physically expected regions (e.g., $\SlopePar \approx 5 \unit{MeV}$).
We leave the resolution of these issues for future work.

Similar improvement as for the \NNLO EDF is observed for the \NLOD EDF. 
This reflects the fact that in $\Delta$-full chiral EFT the dominant two-pion-exchange contribution is promoted from \NNLO to NLO~\cite{Kreb07Deltas}.
At \NNLOD some additional attraction is brought in.
For the interactions used here the additional contributions (which in $\Delta$-less chiral EFT would occur in part at even higher orders) 
are similar in size as the difference between the chiral contributions at \NNLO and \NLOD.
The \gude functionals are generally not sensitive to such differences on a qualitative level; see \cref{tab:rmsd}.

All statements made above dealt with chiral NN interactions only.
The inclusion of 3N forces does not seem to have a significant effect on the description of nuclei and INM at any considered order; see \cref{sec:results}.
In ab initio calculations, 3N forces are important for a quantitative reproduction of nuclei, and are key for shell structure and for the limits of bound nuclei~\cite{Hamm13RMP,Hebe15ARNPS}. 
For instance, for the oxygen isotopes, the additional repulsion from 3N forces moves the location of the predicted neutron drip line in agreement with experiment~\cite{Otsu10Ox,Herg13Ochain,Cipo13Ochain,Jans14Ochain,Stro17ENO}.
In \cref{fig:oxygen_masses} we show the ground-state energies of oxygen isotopes as predicted by a few \gude functionals.
Comparing the \NNLODThreeN results with the other EDFs shows that including 3N forces does not move the location of the neutron drip line for the EDFs. 
Similar conclusions hold for the other \gude variants with 3N forces.
In agreement with other EDF calculations~\cite{Erle12Nature}, all EDFs constructed in this work predict \elm{O}{28} to be the heaviest oxygen isotope stable against emitting two neutrons, while experimentally it is \elm{O}{24}.

The crucial difference between the negligible role of 3N forces observed here and their relevant effects in ab initio calculations lies in the fact that the EDFs constructed here yield good saturation properties also without the presence of chiral 3N forces -- see \cref{tab:Optima,fig:infinite} -- while they are absolutely necessary to achieve reasonable saturation in calculations of INM employing chiral interactions~\cite{Bogn05nuclmat,Hebe11fits}.
In such ab initio calculations, the role of 3N forces is already visible at the HF level, so one could have expected an impact also here.
The fact that this is not the case suggests the fitted EDF terms can compensate missing 3N pion exchanges in the density regime relevant for finite nuclei.

\begin{figure}[tbp]
\includegraphics[width=1\linewidth]{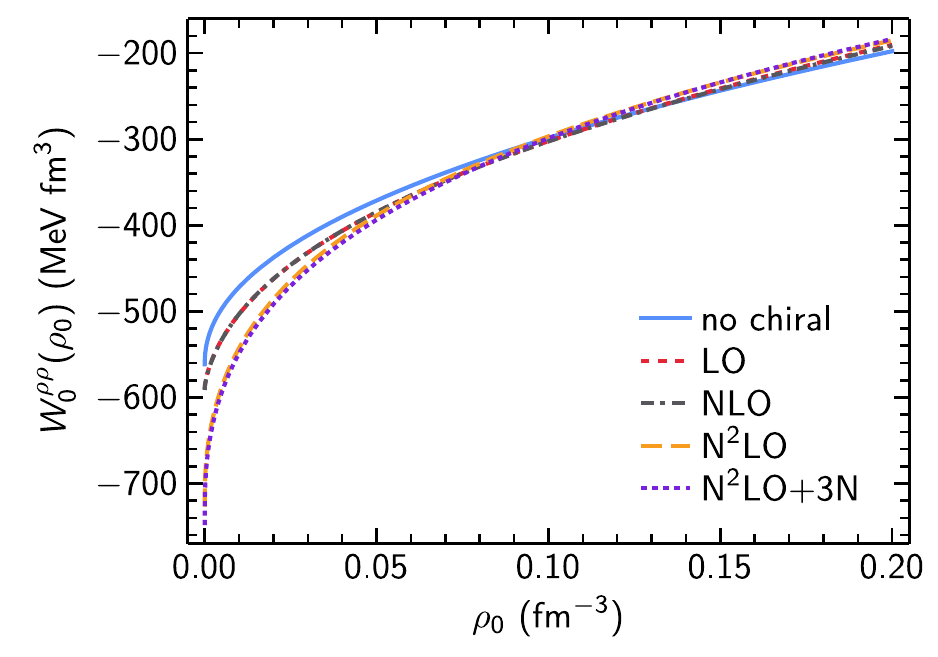}
\caption{\label{fig:rho_total}
$W_0^{\rho\rho}$ for different \gude variants. 
}
\end{figure}

For the terms which depend only on $\rho_0$ this is illustrated in \cref{fig:rho_total}, which shows $W_0^{\rho\rho}$ for different \gude functionals.
The curves for \NNLO with and without 3N forces are basically on top of each other, signaling that for the EDF without 3N pion exchanges the Skyrme part of the EDF mostly takes over the role of the 3N terms (see also the different $\gamma$ values in \cref{tab:Optima}).
This observation correlates well with the original reason to introduce a density-dependent coefficient into nuclear EDFs, namely to replace a genuine 3N interaction~\cite{Vaut71Skyrme}.

The observation that fitting the EDF parameters can almost fully compensate missing 3N pion exchanges is in apparent contradiction with the wrong drip line position observed for the oxygen chain.
In other words the question is, why does the \gude family predict the wrong drip line location even though the functionals either explicitly contain or are essentially able to effectively encapsulate chiral 3N physics?
One simple explanation is the lack of sufficiently neutron-rich nuclei in the experimental data set used in the optimization. 
Since chiral 3N contributions grow with increasing neutron number~\cite{Otsu10Ox,Frim113b},
the description of nuclei closer to stability might not be significantly altered but drip lines might be much improved when optimizing an EDF with chiral 3N contributions using an experimental data set containing more asymmetric nuclei.
Another reason is the importance of beyond-mean-field effects that are known to significantly impact the nuclear structure in light nuclei~\cite{zhou2016anatomy,marevic2019cluster}.

As alluded to above, the existence of strict bounds that we impose on some EDF parameters during their optimization somewhat complicates the analysis of the effect of different chiral contributions.
Some conclusions drawn in the present section might thus not hold in other optimization settings.

\section{\label{sec:SummaryOutlook}Conclusions and outlook}

In this paper, we constructed semi-phenomenological EDFs, dubbed \gude, consisting of pion exchanges taken from chiral EFT at different orders and a phenomenological Skyrme part.
The long-range pion-exchange interactions are included at the Hartree-Fock level (using a DME for the Fock contributions) without adjustment and thereby do not change the number of free EDF parameters.
The \gude functionals with chiral terms perform significantly better than a reference Skryme functional without chiral terms constructed within the same protocol, especially in terms of accurately describing ground-state energies.
These improvements can be traced back to the combination of two terms: Fock contributions from one-pion exchange at leading order in the chiral expansion and Hartree contributions from two-pion exchange at \NNLO.
This is demonstrated with the ``\minchi'' variant of the \gude EDFs which contains only those two terms in addition to the phenomenological part and achieves similar improvements as observed for the other \classTwoHyphen \gude functionals, which contain additional terms stemming from pion exchanges.

Conversely, adding only pion-exchange terms at LO or NLO does not give any improvement. 
While it might seem like a contradiction to the chiral EFT power counting -- according to which the importance of additional terms is reduced with every higher order included -- it may simply result from the fact that we include pion exchanges only at the HF level, i.e., beyond-mean-field effects from pions are not explicitly included and the structure of the contact interactions present in the EDFs does not change with increasing order, unlike in chiral EFT.
Along similar lines, including long-range 3N forces does not yield significant improvement because the optimization procedure of the 
density-dependent 
contact terms in the traditional part of our EDFs allows for the approximate capture of their effects.

The order-by-order systematics of the \gude functionals shows much less variability and surprising behavior compared to what was observed in \rcite{Nava18DMEEDF}, where functionals had been constructed following the same strategy as used here.
In particular, we consider it promising that the inclusion of chiral long-range 3N forces does not lead to a worsening of the EDFs, unlike before.
We attribute this to the different improvements, bugfixes, and other changes established in the present work.
The analysis carried out in \cref{sec:analysis} mostly explains the obtained order-by-order behavior.
In some regards further insight is still needed.
For instance, the detailed mechanism how the improvement is realized at \NNLO (and why some LO terms are additionally needed which on their own do not provide improvement) is still unclear.
We believe that insight might be gained from performing optimizations without imposing bound constraints on INM properties.
It would also be of interest to investigate if adding pion-exchange terms, in particular those included in the ``\minchi'' variant, to other functionals, of Skyrme or other type, gives similar improvement as observed here.
We have also left the study of the dependence of the EDFs on the chiral interactions including their regulators for future work.

Going beyond NLO in the present construction does not only improve the description of finite nuclei, it also considerably changes properties of INM as shown in \cref{tab:Optima}.
The incompressibility $\Incomp$ is significantly reduced and isovector parameters also change strongly.
The decrease of the slope parameter $\SlopePar$ is particularly strong, with it typically ending up at our optimization protocol's lower bound of 30~MeV.

However, in current EDFs isovector terms are generally poorly constrained~\cite{Naza14SymEnDFT,McDo15UncQuaEDF}; the present work is no exception.
This is not of significant consequence when comparing to bulk properties of experimentally accessible nuclei as done here, but limits the predictive power for applications to extreme neutron-rich conditions in astrophysics.
This is because the size of isovector contributions grows significantly when going to very neutron-rich systems.
Including experimental data on neutron skins or dipole polarizabilities~\cite{Piek12DiPolNSk,Naza14SymEnDFT,Rein22MirrorCh} in the optimization the EDF parameters, possibly combined with fitting to ab initio results for neutron drops~\cite{Bogn11DME,Pott14ChiNDrop,Tews16QMCPNM,Bonn18EDFNDrop}, is expected to reduce the uncertainties on the isovector terms.

Extending the optimization data set could also be beneficial in other ways.
Examples are the inclusion of ground-state information for nuclei close to the neutron drip line to better constrain isovector terms and to study the effect of chiral 3N forces, 
and the explicit inclusion of separation energies, which could help with their description and would therefore have significant impact on nucleosynthesis yields from r-process calculations~\cite{Mart16massrpro,Zhu21KiloDep,Gori23NuclRPro}.
All \gude variants underbind nuclei on average.
This might be remedied by increasing the amount of data from open-shell nuclei in the fit or by adjusting the data weights in the optimization.

For practical applications,
correlated uncertainties (or better, distributions) for the EDF parameters should be determined.
They could be estimated using Bayesian inference; see \rscite{McDo15UncQuaEDF,Schu20EDFCalibr,Giul23EDFRBM} for example applications to EDFs.
Such a scheme could also be extended to incorporate expectations for INM parameters via prior distributions in the optimization instead of imposing them as hard parameter bounds as done here.

The GUDE family may be plagued be self-interaction issues~\cite{Bend09PNP}.
For the chiral contributions this is because Fock contributions are included via a DME but the Hartree contributions are included quasiexactly by approximating the chiral potentials as sums of Gaussians.
However, this could be remedied by also treating the Fock terms (at the same chiral order) quasiexactly, which does not lead to significant computational overhead.
In this work, we used the DME because this simplifies the inclusion of 3N forces in EDF frameworks.
However, their inclusion did not lead to significant improvement and they could thus be left out (like in the N$^2$LO GUDE version).
Treating self-pairing effects~\cite{Bend09PNP}, that also occur for conventional functional parametrizations, would require larger adjustments of the EDF structure.

Our work shows that the explicit inclusion of
long-range pion-exchange interactions from chiral EFT at the HF level into a Skyrme EDF improves the description of finite nuclei.
This suggests that such terms will be relevant when generating an EDF completely from first principles.
It might be necessary to account for effects of different types of correlations explicitly to create such an EDF.
Collective correlations may be expected to be captured by going beyond the mean-field description.
However, different instabilities and pathologies occur when EDFs not derived from actual Hamiltonians are used in those frameworks~\cite{Gras19EDFBMF}.
Therefore, functionals of the \gude form could not be used directly.
Partially, these issues would be addressed by incorporating the pion exchanges consistently quasiexactly as discussed above.
Including effects from short-distance correlations from resummed ladder diagrams as described by Brueckner-Hartree-Fock theory should be simpler:
in \rcite{Zhan18BHFDME} density-dependent Skyrme terms generated from a counterterm expansion capturing such correlations were computed.
A next step towards ab initio EDFs could therefore be the inclusion of such terms.

\begin{acknowledgments}

We thank Stefan Wild and Thomas Duguet for helpful discussions and Pierre Arthuis for comments on the manuscript.
In creating some of the figures shown in this paper a colorblind friendly color scheme developed in \rcite{Petr21Color} and an online colorblindness simulation tool \cite{NichColor} were used.
The work of L.Z. and A.S. was supported in part by the Deutsche Forschungsgemeinschaft (DFG, German Research Foundation) -- {Project-ID} 279384907 -- SFB 1245 and by the BMBF Contract No.~05P21RDFNB.
The work of R.J.F. was supported in part by the National Science Foundation under Grants No.~PHY--1913069 and No.~PHY-2209442 and by the NUCLEI SciDAC Collaboration under Department of Energy MSU Subcontract No. RC107839-OSU. 
The work of S.K.B. was supported in part by the National Science Foundation under Grants No.~PHY--1713901 and No.~PHY--2013047, and by the NUCLEI SciDAC Collaboration under Department of Energy Grant No.~de-sc0018083.
Support for this work was partly provided through Scientific Discovery through Advanced Computing (SciDAC) program funded by U.S. Department of Energy, Office of Science, Advanced Scientific Computing Research and Nuclear Physics. It was partly performed under the auspices of the U.S. Department of Energy by the Lawrence Livermore National Laboratory under Contract No. DE-AC52-07NA27344. Computing support for this work came from the Lawrence Livermore National Laboratory (LLNL) Institutional Computing Grand Challenge program.
Calculations for this research were in part conducted on the Lichtenberg high performance computer of TU Darmstadt.
\end{acknowledgments}

\bibliography{literature}

\end{document}